%% file: main.tex
\documentclass[11pt]{article}
\usepackage[top=0.9in, bottom=0.9in, left=0.8in, right=0.8in]{geometry}
\usepackage{setspace}
\usepackage[T1]{fontenc}
\usepackage{fancyvrb}
\usepackage{times}
\usepackage{booktabs}
\usepackage{rotating}
\usepackage{bbm}
\usepackage{dsfont}
\usepackage{graphicx}
\usepackage{multirow}
\usepackage{placeins} 
\usepackage[large, bf]{caption}
\usepackage{subcaption} 
\usepackage{palatino}
\usepackage{pdflscape}
\usepackage{textcomp}
\usepackage{nicefrac}
\usepackage{geometry}
\usepackage{float}
\usepackage{changepage}
\usepackage{makecell}
\usepackage{array}
\usepackage{booktabs} 
\usepackage{threeparttable}
\usepackage{longtable}
\usepackage[hang,flushmargin]{footmisc}
\usepackage[toc,page]{appendix} 
\usepackage{tabularray}
\usepackage{tabularx}
\usepackage{tabu}
\usepackage{capt-of}
\usepackage{wrapfig}
\usepackage{fancyvrb,fvextra} 

\usepackage{abstract}

\makeatletter
\renewcommand{\paragraph}{%
  \@startsection{paragraph}{4}%
  {\z@}{2.2ex \@plus 1ex \@minus .2ex}{-1em}%
  {\normalfont\normalsize\bfseries}%
}
\makeatother

\usepackage{amsmath, amsfonts, amssymb, amsthm}

\usepackage{mathpazo} 
\usepackage{hyperref}
\hypersetup{colorlinks, citecolor=black, filecolor=blue, linkcolor=blue, urlcolor=blue}

\def\urltilda{\kern -.15em\lower .7ex\hbox{\~{}}\kern .04em}

\usepackage{titlesec}


\titlespacing*{\section}{0pt}{0.9ex plus 0.2ex minus .2ex}{0.8ex plus .2ex}
\titlespacing*{\subsection}{0pt}{0.9ex plus 0.2ex minus .2ex}{0.8ex plus .2ex}
\titlespacing*{\subsubsection}{0pt}{1ex plus 1ex minus .2ex}{0.8ex plus .2ex}

\usepackage{xr}

\usepackage{datatool}
\DTLsetseparator{;}
\DTLloaddb[noheader, keys={thekey,thevalue}]{mydata}{outputs/constants.txt} 
\newcommand{\var}[1]{\DTLfetch{mydata}{thekey}{#1}{thevalue}}

\usepackage[
  style=authoryear,
  backend=biber,
  uniquename=false,
  uniquelist=false,
  maxcitenames=1,
  maxbibnames=99, 
  giveninits=true
]{biblatex}

\renewbibmacro{in:}{}
\DeclareNameAlias{author}{family-given}
\DeclareNameAlias{editor}{family-given}
\DeclareFieldFormat[article]{journaltitle}{\mkbibemph{#1}}
\DeclareFieldFormat[article]{volume}{\textbf{#1}}
\DeclareFieldFormat[article]{number}{(#1)}
\DeclareFieldFormat[article]{pages}{#1}

\author{Drew Johnston\textsuperscript{1,*}\quad David Holtz\textsuperscript{2,1}\quad Alex Martin Richmond\textsuperscript{1}\quad\\
Christopher Ong\textsuperscript{1}\quad Prasanna Tambe\textsuperscript{3,1}\quad Aaron Chatterji\textsuperscript{1,4}
\vspace{0.5cm}\\
\textsuperscript{1}OpenAI\quad \textsuperscript{2}Columbia Business School\quad \textsuperscript{3}University of Pennsylvania, Wharton School\\
\textsuperscript{4}Duke University Fuqua School of Business
}

\title{\vspace{-1.7cm}{\textsc{\Huge The Shift to Agentic AI: Evidence from Codex}}}

\date{}
\begin{document}
\begin{filecontents*}{outputs/constants.txt}
consumer_codex_l28_share_pct;0.7
consumer_avg_codex_output_token_share_pct;0.4
consumer_codex_ever_avg_codex_output_token_share_pct;43.8
consumer_aggregate_codex_output_token_share_pct;16.5
enterprise_codex_l28_share_pct;17.3
enterprise_avg_codex_output_token_share_pct;11.7
enterprise_codex_ever_avg_codex_output_token_share_pct;54.0
enterprise_aggregate_codex_output_token_share_pct;63.3
openai_codex_l28_share_pct;97.9
openai_avg_codex_output_token_share_pct;94.1
openai_codex_ever_avg_codex_output_token_share_pct;95.1
openai_aggregate_codex_output_token_share_pct;99.8
internal_job_function_relative_token_usage_baseline_date;November 1, 2025
internal_job_function_min_p50_relative_l28d_combined_output_tokens_fold;10
codeforces_top5_solve_time_classifier_pearson_correlation;0.65
codeforces_rating_classifier_pearson_correlation;0.69
consumer_request_complexity_1h_plus_user_share_dec_2025_pct;35.4
consumer_request_complexity_1h_plus_user_share_may_2026_pct;70.2
consumer_request_complexity_8h_plus_user_share_dec_2025_pct;2.1
consumer_request_complexity_8h_plus_user_share_may_2026_pct;25.6
codex_scim_active_enterprise_firms_count;8629
codex_scim_firms_with_any_coverage_count;3326
codex_scim_firms_with_any_coverage_share_pct;38.5
codex_scim_covered_firm_user_account_coverage_pct;77.6
codex_scim_covered_firm_mean_user_account_coverage_pct;74.9
codex_scim_covered_firm_user_account_coverage_p10_pct;32.8
codex_scim_covered_firm_user_account_coverage_p25_pct;71.2
codex_scim_covered_firm_user_account_coverage_p50_pct;84.2
codex_scim_covered_firm_user_account_coverage_p75_pct;91.1
codex_scim_covered_firm_user_account_coverage_p90_pct;94.8
skill_usage_window_days;7
consumer_skill_any_user_share_pct;25.7
consumer_skill_preinstalled_user_share_pct;11.1
consumer_skill_standalone_curated_user_share_pct;2.2
consumer_skill_plugin_curated_user_share_pct;14.2
consumer_skill_plugin_custom_user_share_pct;9.1
consumer_skill_custom_user_share_pct;11.8
enterprise_skill_any_user_share_pct;30.4
enterprise_skill_preinstalled_user_share_pct;11.1
enterprise_skill_standalone_curated_user_share_pct;2.1
enterprise_skill_plugin_curated_user_share_pct;15.5
enterprise_skill_plugin_custom_user_share_pct;7.8
enterprise_skill_custom_user_share_pct;18.7
openai_skill_any_user_share_pct;96.2
openai_skill_preinstalled_user_share_pct;58.2
openai_skill_standalone_curated_user_share_pct;23.8
openai_skill_plugin_curated_user_share_pct;92.4
openai_skill_plugin_custom_user_share_pct;60.9
openai_skill_custom_user_share_pct;87.2
skill_any_user_share_pct;26.6
skill_preinstalled_user_share_pct;11.2
skill_standalone_curated_user_share_pct;2.2
skill_plugin_curated_user_share_pct;14.5
skill_plugin_custom_user_share_pct;8.9
skill_custom_user_share_pct;13.0
skill_any_user_share_start_pct;5.4
skill_any_user_share_latest_pct;26.6
skill_any_user_share_change_pp;21.2
skill_task_area_application_management_conversation_share_pct;14.9
openai_skill_task_area_application_management_conversation_share_pct;21.0
enterprise_skill_task_area_application_management_conversation_share_pct;14.7
consumer_skill_task_area_application_management_conversation_share_pct;12.1
skill_task_area_code_understanding_conversation_share_pct;15.3
openai_skill_task_area_code_understanding_conversation_share_pct;32.7
enterprise_skill_task_area_code_understanding_conversation_share_pct;16.7
consumer_skill_task_area_code_understanding_conversation_share_pct;12.4
skill_task_area_code_validation_conversation_share_pct;18.4
openai_skill_task_area_code_validation_conversation_share_pct;19.7
enterprise_skill_task_area_code_validation_conversation_share_pct;17.1
consumer_skill_task_area_code_validation_conversation_share_pct;18.2
skill_task_area_collaboration_conversation_share_pct;20.6
openai_skill_task_area_collaboration_conversation_share_pct;50.9
enterprise_skill_task_area_collaboration_conversation_share_pct;20.3
consumer_skill_task_area_collaboration_conversation_share_pct;9.3
skill_task_area_knowledge_artifacts_conversation_share_pct;8.8
openai_skill_task_area_knowledge_artifacts_conversation_share_pct;6.4
enterprise_skill_task_area_knowledge_artifacts_conversation_share_pct;11.9
consumer_skill_task_area_knowledge_artifacts_conversation_share_pct;10.1
skill_subtask_browser_actions_conversation_share_pct;20.1
openai_skill_subtask_browser_actions_conversation_share_pct;27.5
enterprise_skill_subtask_browser_actions_conversation_share_pct;9.6
consumer_skill_subtask_browser_actions_conversation_share_pct;17.5
skill_subtask_code_review_conversation_share_pct;20.8
openai_skill_subtask_code_review_conversation_share_pct;25.8
enterprise_skill_subtask_code_review_conversation_share_pct;17.7
consumer_skill_subtask_code_review_conversation_share_pct;19.9
skill_subtask_meeting_support_conversation_share_pct;25.9
openai_skill_subtask_meeting_support_conversation_share_pct;70.7
enterprise_skill_subtask_meeting_support_conversation_share_pct;22.6
consumer_skill_subtask_meeting_support_conversation_share_pct;5.6
skill_subtask_message_summaries_conversation_share_pct;21.1
openai_skill_subtask_message_summaries_conversation_share_pct;44.0
enterprise_skill_subtask_message_summaries_conversation_share_pct;13.2
consumer_skill_subtask_message_summaries_conversation_share_pct;4.1
skill_subtask_presentations_conversation_share_pct;28.6
openai_skill_subtask_presentations_conversation_share_pct;44.8
enterprise_skill_subtask_presentations_conversation_share_pct;25.5
consumer_skill_subtask_presentations_conversation_share_pct;21.9
skill_subtask_repo_management_conversation_share_pct;20.2
openai_skill_subtask_repo_management_conversation_share_pct;28.7
enterprise_skill_subtask_repo_management_conversation_share_pct;21.4
consumer_skill_subtask_repo_management_conversation_share_pct;16.9
skill_subtask_spreadsheets_conversation_share_pct;17.2
openai_skill_subtask_spreadsheets_conversation_share_pct;52.2
enterprise_skill_subtask_spreadsheets_conversation_share_pct;4.0
consumer_skill_subtask_spreadsheets_conversation_share_pct;20.5
skill_subtask_task_management_conversation_share_pct;21.6
openai_skill_subtask_task_management_conversation_share_pct;43.0
enterprise_skill_subtask_task_management_conversation_share_pct;27.8
consumer_skill_subtask_task_management_conversation_share_pct;18.0
persona_developer_user_share_pct;61.0
persona_general_knowledge_worker_user_share_pct;29.3
persona_personal_user_share_pct;9.8
persona_openai_developer_user_share_pct;58
persona_enterprise_developer_user_share_pct;67
persona_consumer_developer_user_share_pct;59
openai_daily_task_runtime_p50_hours_end_date;2.5
openai_daily_task_runtime_p99_hours_end_date;71
openai_daily_task_runtime_p99_growth_since_start_pct;88
task_runtime_daily_series_start_date;April 7, 2026
chatgpt_tool_turn_share_pct;21.9
codex_tool_turn_share_pct;60.3
codex_to_chatgpt_tool_turn_share_ratio;2.75
codex_minus_chatgpt_tool_turn_share_pp;38.4
turn_concurrency_start_date;2026-06-05
turn_concurrency_end_date;2026-06-11
consumer_turn_concurrency_peak_1_user_share_pct;63.9
enterprise_turn_concurrency_peak_1_user_share_pct;67.4
openai_turn_concurrency_peak_1_user_share_pct;10.7
openai_turn_concurrency_peak_5plus_user_share_pct;28.6
codex_analysis_end_date;June 11, 2026
codex_scim_engineering_technical_practitioner_avg_user_codex_output_token_share_pct;26.8
codex_scim_engineering_technical_practitioner_aggregate_codex_output_token_share_pct;88.3
codex_scim_data_analytics_practitioner_avg_user_codex_output_token_share_pct;15.2
codex_scim_data_analytics_practitioner_aggregate_codex_output_token_share_pct;71.9
codex_scim_people_recruiting_role_avg_user_codex_output_token_share_pct;2.0
codex_scim_people_recruiting_role_aggregate_codex_output_token_share_pct;13.8
codex_scim_legal_role_avg_user_codex_output_token_share_pct;1.9
codex_scim_legal_role_aggregate_codex_output_token_share_pct;17.6
internal_job_function_legal_p50_relative_l28d_combined_output_tokens_fold;13
internal_job_function_research_p50_relative_l28d_combined_output_tokens_fold;56
consumer_request_complexity_8h_plus_user_share_dec_2025_to_may_2026_growth_fold;12.3
turn_concurrency_3plus_user_share_pct;11.7
\end{filecontents*}

\maketitle
\begingroup
\renewcommand\thefootnote{}
\footnotetext{The authors are grateful to Cassandra Duchan Solis for her substantial contributions to this paper, and to Tejal Patwardhan, Suproteem Sarkar, and Katy Shi for helpful early conversations and feedback. Andrew Hillis and Kevin Liu provided helpful insights about the data infrastructure underlying many of the analyses in this paper. Gawesha Weeratunga arranged the data access supporting our analyses of usage within organizations. Daniella Eguiguren and Kirsty Le produced the web visualization, which was developed by Matt Nichols, Shawn O'Mara, and Dion Peters. Charlene Tsang designed portions of the visualization. Rachel Brown, Laurance Fauconnet, Rob Friedlander, and Laurie Jones provided editorial feedback. Laura Bisesto, Sarah Friar, Zak Weiler, and Meng Jia Yang reviewed this paper. David Holtz and Prasanna Tambe contributed to this work in their capacity as paid contractors for OpenAI.\\
\textsuperscript{*}Corresponding author: \href{mailto:drewjohnston@openai.com}{drewjohnston@openai.com}
}
\endgroup

\begin{center}
\begin{minipage}{0.9\textwidth}
\setstretch{1.05}
\small
\hspace{-0.3cm}\begin{abstract}
\noindent We analyze usage data from OpenAI's Codex tool to present large-scale evidence of how agentic AI technology, which can take actions on a user's behalf, changes how people work. We use an automated, privacy-protecting pipeline to contrast usage across three populations: external personal-account users, external organizational-account users, and workers within OpenAI. We find that agentic AI usage is growing rapidly: the number of active users has grown more than fivefold in the first half of 2026, with the most rapid increase occurring outside the initial audience of software developers. Uptake is uneven across contexts: within OpenAI, Codex usage is nearly universal and has largely replaced business usage of ChatGPT. We document a similar shift to agentic tooling outside OpenAI, particularly within organizations, although external adoption remains lower and more uneven. In addition to headline usage figures, we observe measures of sophistication, and find that a growing number of users have used Codex to change their workflows substantially. We find that more than 10\% of users manage three or more concurrent Codex agents at some point each week and that \var{skill_any_user_share_pct}\% use skills, which allow users to share instructions for complex workflows. Alongside these changes in usage practices, request complexity has increased: since the start of the year, the share of individual Codex users who submit at least one request for a task estimated to require more than eight hours for an experienced human to complete has increased nearly tenfold. Concurrently, output has grown rapidly---in June 2026, the median OpenAI employee in a legal role generated 13 times more monthly output tokens across Codex and ChatGPT than they did in November 2025, while the median researcher generated more than 50 times as many. We conclude by discussing the implications of these patterns for productivity, job reorganization, and workforce restructuring.

\end{abstract}
\end{minipage}
\end{center}

\vfill  
\pagebreak
\section{Introduction}

Generative AI systems increasingly differ in the extent to which they can act on a user's behalf, or in other words, the extent to which they are \textit{agentic}. Traditional chatbot interfaces are primarily conversational: users ask questions or provide instructions, and the model generates responses. Agentic AI tools extend this paradigm by enabling users to delegate multi-step tasks to AI systems that can autonomously use external tools, inspect files, execute commands, and create or modify artifacts.

This paper studies this shift to agentic AI by analyzing data on the adoption and use of Codex, OpenAI's agentic\footnote{In this paper, for brevity, we refer to ChatGPT as a conversational AI tool and Codex as an agentic AI tool. In practice, the distinction is not absolute. ChatGPT includes agentic capabilities, such as web browsing and code execution, while some Codex interactions are purely conversational. In the week preceding June 11, 2026, \var{codex_tool_turn_share_pct}\% of Codex turns and \var{chatgpt_tool_turn_share_pct}\% of ChatGPT turns invoked at least one external tool. Tool use is an imperfect proxy for agency: some tool invocations are part of simple conversational interactions, while some agentic workflows involve limited tool use.} coding and work platform, which was initially released in April 2025.\footnote{This initial release was limited in scope and available only via a command line interface. Later releases broadened the scope of users eligible to use Codex and the variety of surfaces through which it could be accessed.} Codex was originally designed for software development, a domain where AI outputs are useful, economically important, and comparatively easy to verify. However, the usage we study extends beyond software: Codex is used for drafting documents, analyzing data, coordinating communication, and other knowledge-work tasks outside coding.

To study how agentic AI use diffuses across settings and expands across task domains, we analyze Codex usage data. These data allow us to document adoption, task allocation, and depth of use, using automated classifiers to extract aggregated and anonymized insights about AI usage without researchers reading the underlying messages themselves.

We conduct this analysis across three distinct populations: individual users, organizational users, and OpenAI workers.\footnote{We refer to users on personal plans, including Free, Go, Plus, and Pro, as Individual users. We refer to users on Business and Enterprise plans as Organizational users.} Individual users provide a broad-market view of early adoption, where use is heterogeneous and workflow integration is often limited. Organizational users show how Codex is used in business settings outside OpenAI, where adoption may depend on local workflows, permissions, security requirements, and user familiarity. OpenAI workers provide a unique view into what usage is at the frontier. OpenAI is an unusually favorable environment for agentic AI: workers are highly familiar with frontier models, usage is cheap at the margin, organizational buy-in is high, training and informal knowledge sharing are common, and many workflows are close to the systems being developed.\footnote{Internally at OpenAI, there are no quantity restrictions on AI usage, and there is significant internal knowledge sharing about AI capabilities, skills, and usage patterns.} OpenAI usage is therefore not representative of the typical organization today. However, it provides a view of what agentic AI use may look like in the future, when adoption frictions are minimal.

We organize the evidence around four stylized facts.

First, the shift to agentic AI is rapid but uneven. Over the first six months of 2026, weekly active Codex usage increased sharply, yet agentic tooling remains much less broadly used than ChatGPT. The shift is smallest among individual users, larger among organizational users, and largest among OpenAI workers. The contrast is sharpest when measured through the use of output tokens rather than numbers of active users.\footnote{Tokens are the basic unit of information read and generated by most AI models.} Among OpenAI workers, Codex largely replaced ChatGPT as the main interface for work-related AI use: as of \var{codex_analysis_end_date}, Codex accounts for \var{openai_aggregate_codex_output_token_share_pct}\% of output tokens these workers generate across Codex and ChatGPT. Among organizational users, the corresponding share is \var{enterprise_aggregate_codex_output_token_share_pct}\%, while among individual users it is \var{consumer_aggregate_codex_output_token_share_pct}\%. Adoption also varies by job role. Technical roles adopt earlier, but non-developer use has grown quickly, especially within OpenAI.

Second, Codex use is strongly oriented toward delegated production. Users ask the model to carry out concrete work tasks like debugging, refactoring, validating changes, configuring applications, drafting documents, and analyzing data. These activities are better understood as production than as consultation: users are asking Codex to do work, not only to provide advice or information.\footnote{This contrasts with conversational AI usage patterns documented in \textcite{chatterji2025people}. In that context, researchers found that a larger share of work focused on ``asking'' (which represented nearly half of all prompts), than on ``doing''.} This distinction is central to the shift we study. In an agentic interface, usage can correspond to a delegated workflow rather than a single conversational exchange. We find that the complexity of tasks users request has increased over time, with an increasing share of users delegating tasks that would take an experienced human more than a full day of work to complete.

Third, Codex use is anchored in software production, but is broader in contexts in which adoption is deepest. Across user populations, the largest share of tasks is closely tied to software work, including code implementation, code understanding, code validation, engineering operations, and application management. But users delegate more than code generation, using Codex across the broader software-production life cycle: they also use it to understand existing systems, configure environments, validate changes, manage repositories, and produce documentation. Among OpenAI workers, usage extends further into research, planning, communication, data analysis, product work, recruiting, sales, and other non-engineering activities.

Fourth, intensive users organize Codex use around large, repeatable, and parallel workflows. We measure this in several ways. Output volume measures the amount of AI-mediated work users generate. Task-complexity measures the estimated time it would take an experienced human to complete the tasks that users delegate. Skill use measures whether users invoke reusable instructions, capabilities, or tool integrations. Runtime measures how much active agent work occurs on a user's behalf. Concurrency measures whether users run multiple Codex turns at the same time. Across these measures, heavy users look qualitatively different from occasional users. They are more likely to use skills, run longer and more complex tasks, and operate multiple agents concurrently. OpenAI workers provide the clearest view of this pattern: among the most intensive users within OpenAI, Codex is less an assistant answering requests and more like a workflow system in which the user delegates, monitors, reviews, and coordinates multiple streams of work.

The evidence illustrates why agentic AI is more than a more capable version of conversational AI: users can hand off larger pieces of work to systems that inspect context, use tools, and modify artifacts. This distinction changes what adoption means. For agentic AI, the important margins are not only whether a person uses the tool, but what work they delegate, how much execution the system performs, and whether users begin to organize workflows around repeated or parallel delegation.

The rest of the paper proceeds as follows.  Section \ref{sec:lit} describes the related literature and connects it to the work of this paper. Section~\ref{sec:who} studies who is using agentic tooling, documenting growth in Codex usage across users in individual, organizational, and internal OpenAI accounts. Within these categories, we study differences in usage across personas, job functions, and seniority levels. Section~\ref{sec:tasks} examines what people use Codex for, comparing task composition across user populations and worker groups. Section~\ref{sec:how} studies how people use Codex once they adopt it, focusing on output volume, repeatable workflows, task complexity, runtime, and concurrency. The conclusion summarizes what these usage patterns show about the frontier of agentic AI use.

\section{Related literature}
\label{sec:lit}

In studying the diffusion and use of Codex, we connect the emerging literature on agentic AI to three related literatures that study generative AI adoption, delegation and verification in human--AI systems, and the organizational complements that shape technology diffusion.

First, we extend work on the adoption and use of generative AI. Prior work measures occupational exposure to large language models, generative AI adoption by workers, and usage of conversational AI systems \parencite{eloundou2024gpts,felten2023occupational,bick2024rapid,chatterji2025people,handa2025economic}. This literature documents the rapid diffusion of generative AI across occupations and industries while highlighting substantial heterogeneity in adoption and use. Recent studies also describe how individuals employ conversational AI systems in practice, finding that information retrieval, learning, writing, content creation, and practical guidance account for a large share of observed usage \parencite{chatterji2025people}.

We extend this work by studying agentic AI tools, where the relevant unit of analysis is a delegated workflow rather than a conversation. Like other papers in this literature, our goal is to understand the underlying tasks people perform with new technologies by analyzing interactions between users and AI systems. However, the agentic nature of Codex allows us to move beyond understanding conversations and to directly observe actions performed on a user's behalf.

This distinction connects our work to an emerging literature on agentic AI tools. The closest papers to our work study usage of Perplexity's agentic products. \textcite{yang2025perplexity} study the adoption of a general-purpose browser agent and show that early agentic use is broad, spanning productivity, learning, shopping, media, career, and travel tasks, with usage varying across personal, professional, and educational contexts. \textcite{yang2026reshape} compare Perplexity's Search and Computer products using matched tasks, showing that the more autonomous Computer product performs substantially more work on users' behalf, reduces completion time, and expands the scope of tasks users attempt. A related software-focused literature studies coding agents in practice. \textcite{sarkar2025agents} uses data from an AI-assisted programming platform to show that agents shift software work away from implementation activities such as typing code and toward delegation. \textcite{baumann2026swechat} introduce SWE-chat, a dataset of real coding-agent sessions from open-source developers, and show that coding-agent use in the wild is heterogeneous, involves substantial user correction, and often differs from benchmark-style evaluations. More broadly, a growing literature examines how agentic systems reshape the allocation of work between humans and machines, creating new opportunities for delegation while introducing new requirements for supervision, verification, and coordination.

These questions connect the emerging literature on agentic AI to a long tradition of research on technology diffusion, organizational complements, and workplace change. A central insight from this literature is that productivity gains from new technologies depend on complementary investments in business processes, worker skills, organizational design, and intangible capital. For example, \textcite{brynjolfsson2017productivity} argue that the gap between rapid AI progress and slow measured productivity growth reflects the time required for organizations to develop complementary innovations and organizational changes.

A shift toward agentic AI changes not only what tasks can be automated but also how work is organized. \textcite{hitzig2026agentic} argue that agentic systems move human--AI interaction from assistance toward delegation, making supervision, verification, and coordination central determinants of value creation while increasing returns to domain expertise.  More broadly, realizing value from agentic AI may depend on the diffusion of AI capabilities throughout the workforce and the growing importance of domain expertise in supervising and coordinating delegated work \parencite{tambe2026reskilling,hitzig2026agentic}. Consistent with this view, \textcite{demirer2026writing} show that large task-level productivity gains may translate only imperfectly into final output because downstream human activities remain important bottlenecks. Similarly, \textcite{demirer2026chaining} find that AI generates the largest productivity gains when it can execute contiguous chains of activities, highlighting the importance of workflow structure and task adjacency for automation and job redesign.

Recent workplace evidence reinforces these arguments. \textcite{dillon2025shifting} study a field experiment in which workers received access to an integrated generative AI tool for email, meetings, and writing. They find meaningful individual-level time savings, especially in email, but limited evidence that access changed the quantity or composition of workers' tasks. Similarly, \textcite{chen2025firm} show that the labor-market effects of generative AI depend on whether occupations are more exposed to automation or augmentation. Using job-posting data, they find reduced demand and skill requirements in automation-prone jobs but increased demand and skill complexity in jobs involving human--AI collaboration. Together, these findings suggest that realizing the value of AI depends not only on technological capability but also on how organizations redesign workflows, responsibilities, and skill requirements around new forms of human--AI collaboration.

Our evidence on differences across Individual, Organizational, and OpenAI users speaks directly to these organizational-complement mechanisms. If agentic AI adoption depended only on model capability, we would expect similar patterns of use wherever the same model is available. Large differences across user groups instead suggest that adoption depends on context: access to relevant files and systems, management expectations, workforce skills, and the availability of complementary review processes. In this sense, agentic AI is not simply a cheaper input into existing work; its value depends on whether organizations can redesign workflows, responsibilities, and review processes around delegation and verification.

\section{Who Uses Agentic AI?}
\label{sec:who}

We begin by studying who uses Codex, and how adoption of agentic tooling differs across user populations and worker groups. 
Codex usage grew rapidly after launch, with the number of weekly active users increasing more than fivefold between January 1 and June 1, 2026. This establishes the aggregate growth of Codex, but user counts alone do not capture the shift from conversational to agentic tooling. Many users continue to use ChatGPT, and users who adopt Codex often use it much more intensively than users who remain only on conversational interfaces. For that reason, after documenting aggregate weekly active usage, we focus primarily on Codex's share of output tokens across Codex and ChatGPT.

\begin{figure}[ht!]
    \centering
    \caption{\raggedright Codex usage relative to ChatGPT, by user population}
    \label{fig:codex_chatgpt_relative_usage}
\begin{subfigure}[t]{\linewidth}
    \centering
        \includegraphics[width=\linewidth]{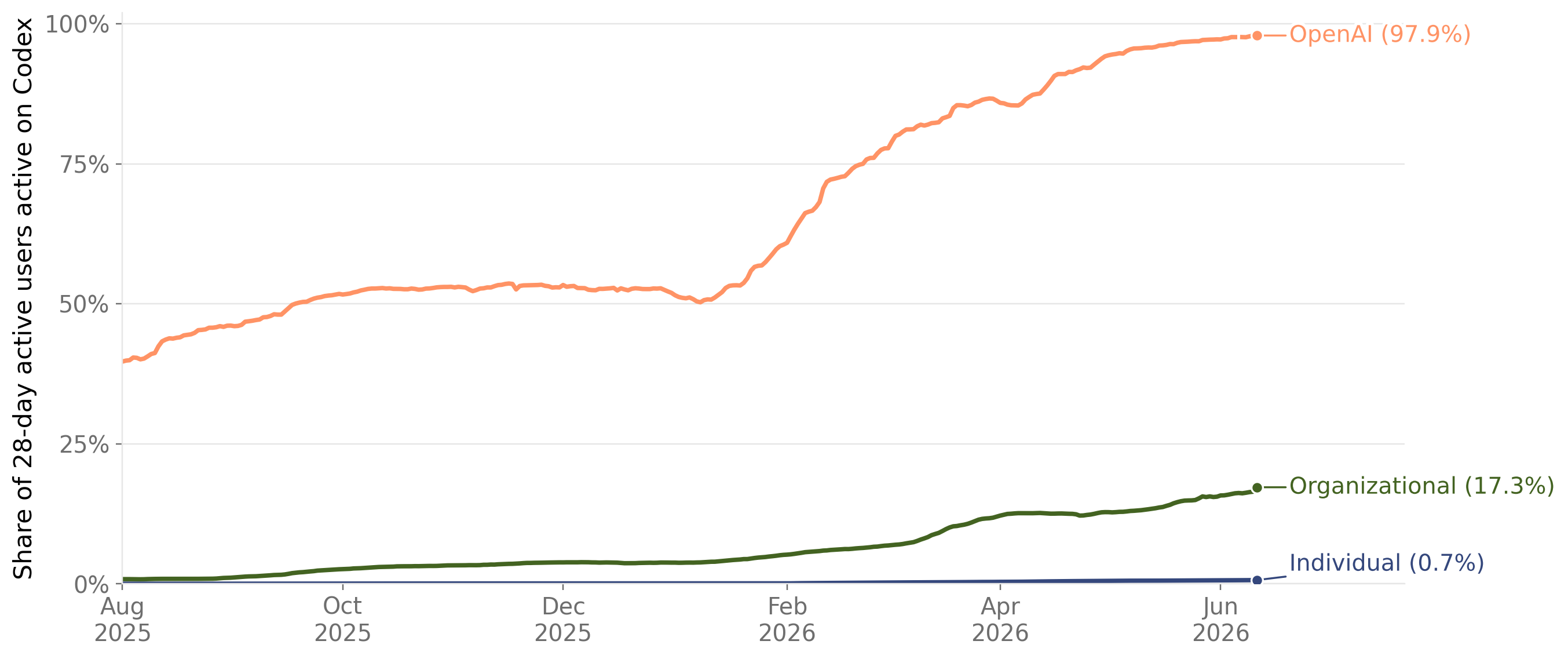}
    \caption{\raggedright Share of active users using Codex}
    \label{fig:codex_wau_share}
\end{subfigure}
\vspace{0.5cm}
\begin{subfigure}[t]{\linewidth}
    \centering
        \includegraphics[width=\linewidth]{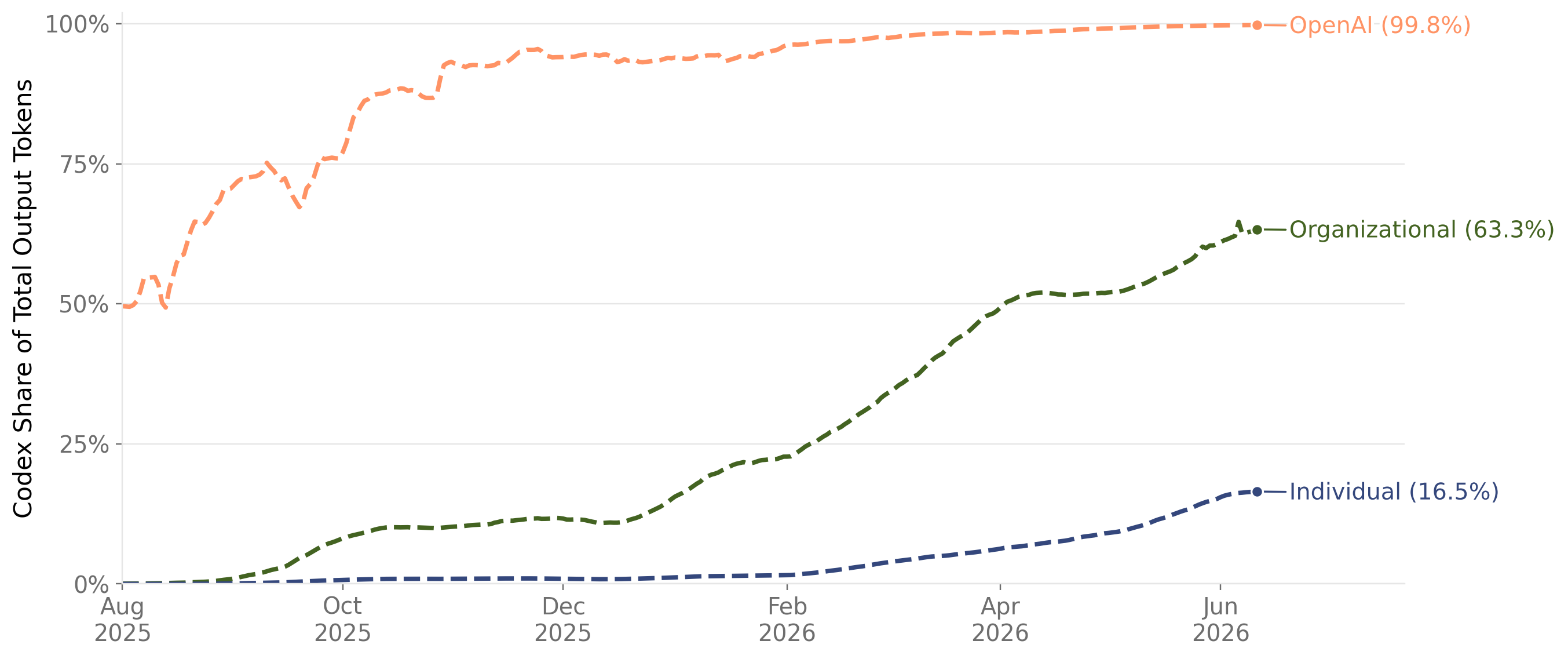}
        \caption{\raggedright Share of output tokens from Codex}
    \label{fig:codex_token_share}
\end{subfigure}
    \begin{minipage}{\linewidth}
    \vspace{-0.5cm}
    \footnotesize{Panel A shows the share of users active on either ChatGPT or Codex during the preceding twenty-eight days who were active on Codex at least once during that period. Panel B shows output tokens produced on Codex as a share of output tokens produced on either Codex or ChatGPT during the preceding twenty-eight days. We compare individual users, organizational users, and OpenAI workers.}
    \end{minipage}
\end{figure}

Figure~\ref{fig:codex_chatgpt_relative_usage} compares the shift to Codex across OpenAI workers, organizational users, and individual users. Panel A shows the extensive margin: whether active users of either product use Codex at all. Panel B shows the intensive margin: the share of output tokens produced through Codex rather than ChatGPT. Among individual users, conversational interfaces continue to dominate: fewer than 1\% of active individual users used Codex in the last 28 days. But the individual users who adopt Codex are unusually intensive, so Codex accounts for a much larger share of individual-user output tokens than of individual active users. Among organizational users, adoption is broader. In the last 28 days, \var{enterprise_codex_l28_share_pct}\% of organizational users used Codex, several times the corresponding share among individual users, and Codex accounts for the majority of output tokens generated by organizational users in recent weeks. The most pronounced shift is among OpenAI workers. Codex use is both pervasive and intensive: almost all active OpenAI workers use Codex each week, and in recent weeks Codex accounts for more than 99\% of output tokens generated across Codex and ChatGPT.

\begin{figure}[thb]
    \centering
    \caption{\raggedright Growth in active Codex users by account type and persona}
    \includegraphics[width=\linewidth]{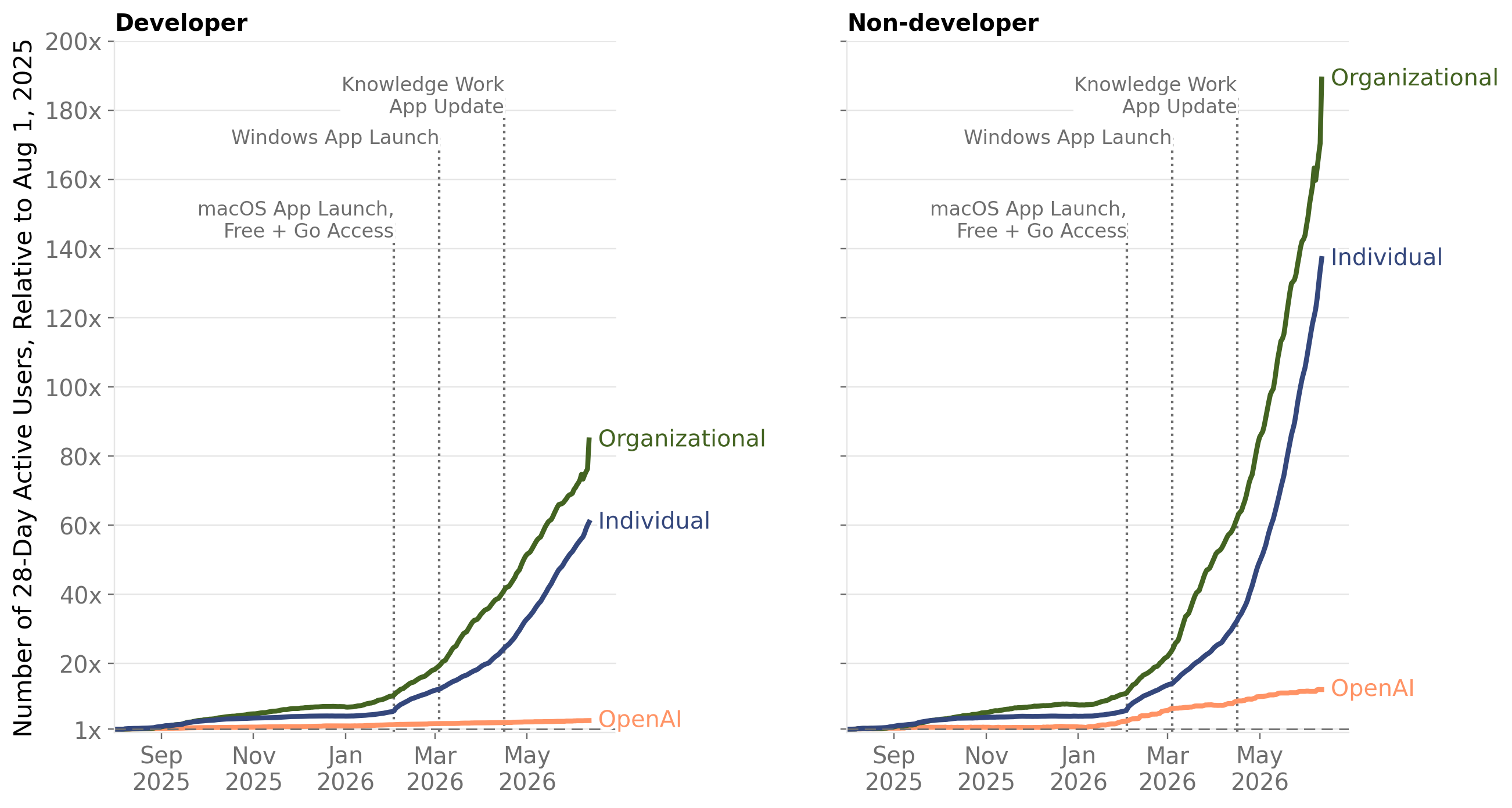}
    \label{fig:persona_share}
    \begin{minipage}{\linewidth}
    \footnotesize{Figure shows growth in users active in the previous 28 days, by persona and account type. The Non-developer series pools users assigned to the Personal and General Knowledge Worker personas. Personas are assigned using the most common label assigned to a user's queries in the 30 days prior to \var{codex_analysis_end_date}. Users' account types are allowed to vary over time. All series are indexed to have a value of 1 as of August 1, 2025. Estimates are based on a 3\% sample of users.}
    \end{minipage}
\end{figure}

To assess whether Codex's growth is driven mainly by developers or by a broader set of users, we classify Codex users into three broad personas based on their recent Codex requests: \emph{Developers}, \emph{General Knowledge Workers}, and \emph{Personal} users. Each request receives a persona label from an automated classifier, and each user is assigned the persona that appears most often in their Codex activity over the prior 30 days, with ties broken by the most recent request. The prompt used for this classification is provided in Appendix~\ref{app:persona_prompt}.\footnote{We validated the persona classifier using a small sample of employees. For each employee, we compared the assigned persona against their HR job title to assess alignment. The results showed strong agreement: over 90\% of engineers were classified into the Developer persona, while more than 90\% of sales employees were classified into the General Knowledge Worker persona.}

Software development activity is assigned the Developer label, other work activity is labeled as General Knowledge Worker, and non-work usage is combined into the Personal bucket. Typical Developer use includes writing, reviewing, or refactoring code. General Knowledge Worker use includes writing or editing documents, analyzing spreadsheets, drafting memos, coordinating communication, and other non-coding tasks. Personal users tend to use Codex for tasks including hobbies, financial activities, and education.

Figure~\ref{fig:persona_share} shows that Codex adoption is not limited to the initial developer base. Developers remain an important share of users, especially among individual and organizational accounts, but growth is faster among Non-developers. Appendix Figure~\ref{fig:persona_detailed_share} provides a more detailed decomposition of personas for individual and organizational users. It shows that, in contrast to use within OpenAI, individual and organizational use remains more concentrated in developer personas, especially full-stack engineering.

\begin{figure}[!t]
    \centering
\caption{\raggedright Codex share of output tokens for the average user, by worker type}
\label{fig:codex_token_worker_type}
\begin{subfigure}[t]{0.9\linewidth}
    \centering
        \includegraphics[width=\linewidth]{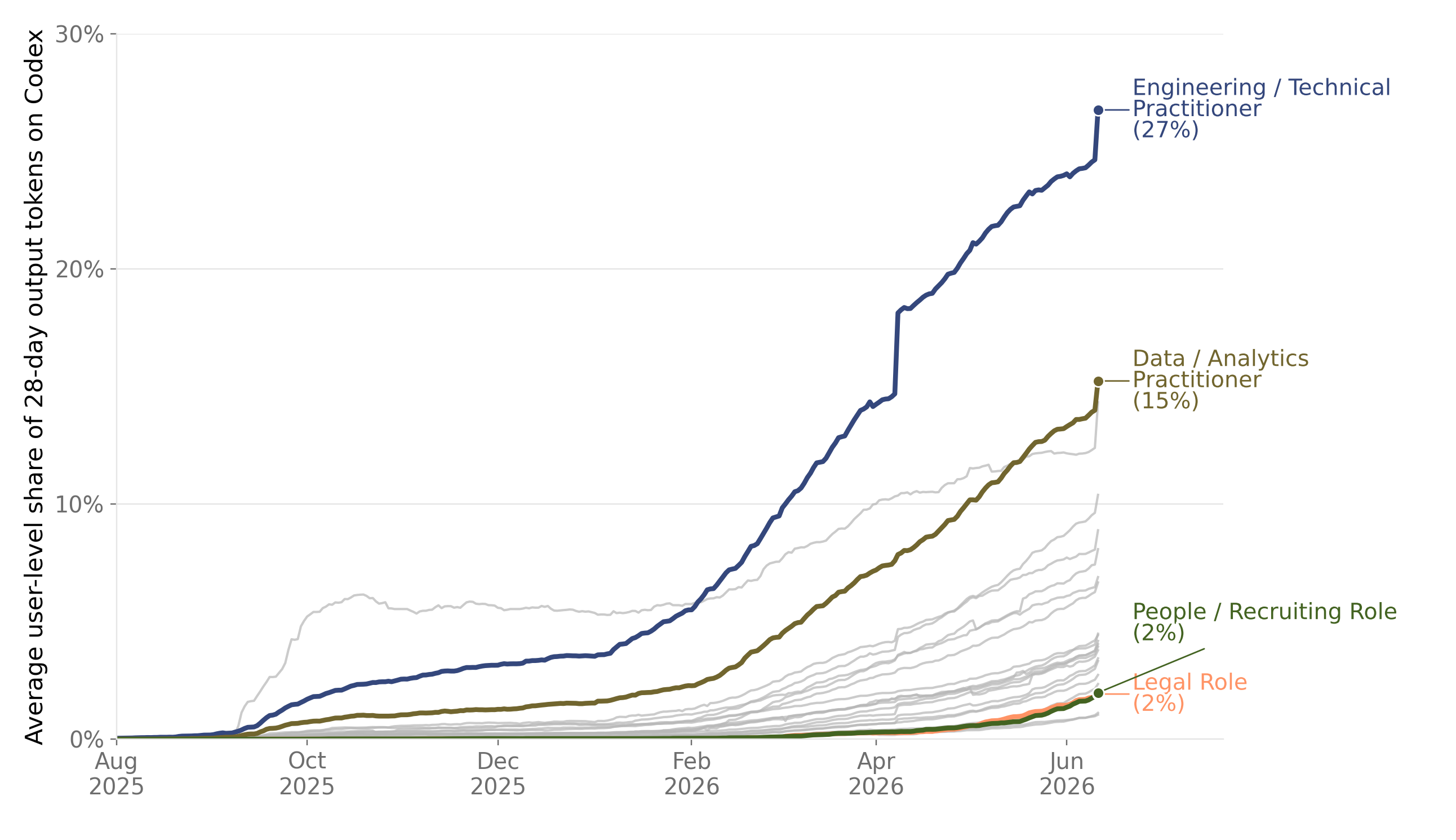}
    \caption{\raggedright Organizational users, by inferred job title}
    \label{fig:codex_token_title}
\end{subfigure}
\vspace{0.5cm}
\begin{subfigure}[t]{0.9\linewidth}
    \centering
        \includegraphics[width=\linewidth]{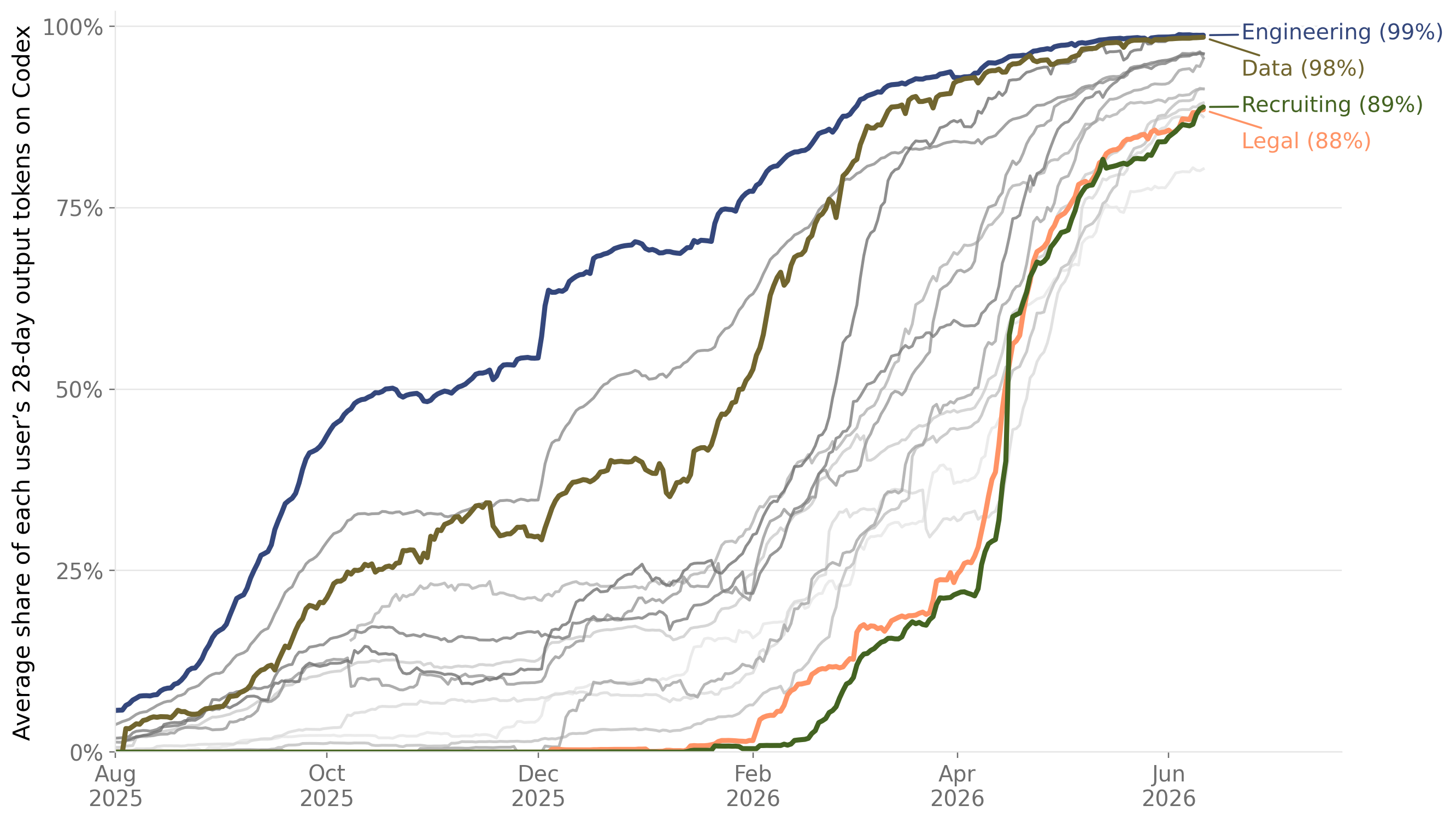}
        \caption{\raggedright OpenAI workers, by job function}
    \label{fig:openai_usage_by_department}
\end{subfigure}
    \begin{minipage}{\linewidth}
    \vspace{-0.5cm}
    \footnotesize{Panel A shows the average user-level share of output tokens across ChatGPT and Codex that were produced on Codex among organizational users in firms with high-quality job-title information, by inferred job-title class. Panel B shows the corresponding employee-level share among OpenAI workers, by job function. Figures are based on output tokens generated on each platform in the preceding 28 days. Only users active in the preceding 28 days are included.}
    \end{minipage}
\end{figure}

We next study adoption by job title and function, moving from behavior-based personas to organizational role measures. For organizational users outside OpenAI, we focus on the subset of firms for which we have high-quality job-title information.\footnote{Within this set of firms, job-title coverage is broad: for the median firm, we observe job titles for \var{codex_scim_covered_firm_user_account_coverage_p50_pct}\% of workers.} We classify job titles into cleaned job-title groups, seniority levels, and people manager statuses using the classifier described in Appendix~\ref{app:job_title_classification}. For OpenAI workers in this analysis, we instead use directly observed internal job-function metadata.\footnote{We use classifier-based measures for OpenAI users elsewhere in the paper. In this analysis, however, internal job-function metadata provides a higher-quality role measure, so we use it instead of classifier-derived job-title groups.} These measures allow us to study whether the shift from ChatGPT to Codex is concentrated in technical roles or whether it extends across the organization.

Figure~\ref{fig:codex_token_worker_type} highlights broad patterns of adoption. Across all job functions, the share of output tokens generated on Codex is increasing, with the most rapid progress among technical job functions. Among Organizational account users, the average engineer produces \var{codex_scim_engineering_technical_practitioner_avg_user_codex_output_token_share_pct}\% of their tokens on Codex, a share that has quintupled since the start of the year. Other technical roles, such as Data and Analytics Practitioners, have seen similarly rapid increases, with the average user in such a role now generating about \var{codex_scim_data_analytics_practitioner_avg_user_codex_output_token_share_pct}\% of their output tokens on Codex. Though non-technical roles have seen a large proportional increase, the average user in a non-technical role generates a relatively small share of their tokens in Codex, with Codex usage among legal teams accounting for only \var{codex_scim_legal_role_avg_user_codex_output_token_share_pct}\% of output tokens for the average user.

Among Organizational account users, the relatively low share of tokens generated across job functions is primarily due to the limited share of users who use Codex at all, which we highlight in Figure~\ref{fig:codex_chatgpt_relative_usage}. Despite the fact that Codex usage has not diffused fully in most job functions, it nevertheless accounts for a large share of overall tokens generated among many job functions, since usage intensity is much higher among users who adopt Codex. For instance, although Codex accounts for just \var{codex_scim_engineering_technical_practitioner_avg_user_codex_output_token_share_pct}\% of the output tokens generated by the average engineer with an Organizational account, Codex usage accounts for \var{codex_scim_engineering_technical_practitioner_aggregate_codex_output_token_share_pct}\% of the total output tokens generated by users of this type. This reflects both its adoption among more intensive users within each function and the deeper use cases agentic tooling enables. We see a similar divide in non-technical roles: among legal users, Codex accounts for \var{codex_scim_legal_role_aggregate_codex_output_token_share_pct}\% of total output tokens, despite accounting for only \var{codex_scim_legal_role_avg_user_codex_output_token_share_pct}\% of tokens generated by the average user. 

Usage within OpenAI provides a glimpse into a similar transition where workers began to shift usage to Codex from conversational harnesses much earlier. In the second half of 2025, usage patterns within OpenAI looked quite similar to those in external organizational accounts today; engineers used agentic harnesses for the largest share of their usage, while usage in other job functions lagged behind. Over the following months, as tooling improved and Codex usage diffused within the company, Codex continued to rise in all departments, with usage accounting for more than 90\% of the average engineer's usage by March 2026, and for more than 90\% of use in almost all departments by the present day. 

The transition to agentic tooling was more rapid in later-adopting job functions within OpenAI than it was among the early-adopting technical functions. In January 2026, when engineering users had already shifted more than half of their AI usage to Codex, use in functions such as legal and recruiting was close to zero. From that point, usage increased gradually to about 20\% of output tokens by early April, then rose quickly from roughly 20\% to 75\% within a month. Although differences across functions remain, convergence toward agentic tooling was rapid: between December 2025 and April 2026, OpenAI moved from a pattern in which most functions primarily used conversational AI to one in which Codex was dominant across functions. This period coincided with a broader internal effort to adopt Codex across workflows, including training sessions and regular feedback loops.

\begin{figure}[!t]
    \centering
\caption{\raggedright Codex share of output tokens among Organizational users, by inferred seniority level}
\label{fig:codex_token_seniority}
        \includegraphics[width=\linewidth]{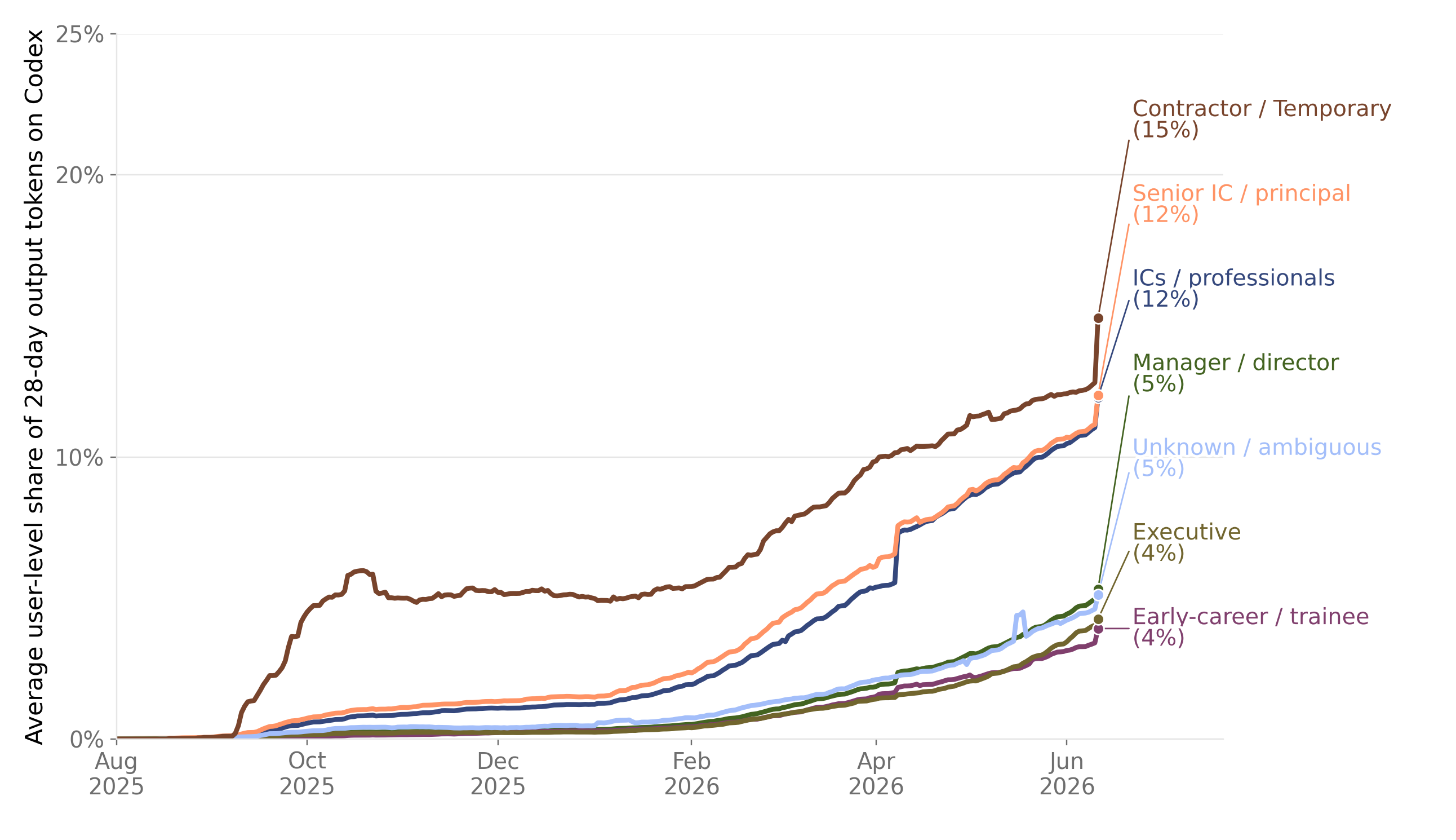}
    \begin{minipage}{\linewidth}
    \footnotesize{Each series shows the average user-level share of output tokens across ChatGPT and Codex that were produced on Codex for organizational users with high-quality job-title information, by inferred seniority level. Figures are based on output tokens generated on each platform in the preceding 28 days. Only users active in the preceding 28 days are included.}
    \end{minipage}
\end{figure}

Finally, Figure~\ref{fig:codex_token_seniority} shows that adoption among Organizational users is not confined to a single seniority level. Codex's share of output tokens rises across the seniority distribution, indicating that agentic tooling is used by both more junior and more senior workers. This result matters because Codex is not only a tool for direct implementation by individual contributors. Senior users may also use agentic tooling for planning, review, and other work that involves delegating tasks and evaluating outputs. We return to these differences in delegated work in Section~\ref{sec:tasks}.

The evidence in this section shows where the shift to Codex is occurring: adoption is broadest and most intensive among organizational users and OpenAI workers, it begins among technical workers before spreading to non-developer roles, and is occurring across all seniority levels within organizations. The next section asks what these users do with Codex once they adopt it. That task-level evidence is important because account type, persona, and job function do not by themselves reveal whether Codex is being used for software development, broader knowledge work, or personal tasks.

\section{What kinds of work do people do with Agentic AI?}
\label{sec:tasks}

This section studies what users delegate to Codex once they adopt it. We use automated classifiers to measure two aspects of delegated work: the type of task being attempted and, for a smaller sample, the estimated human time required to complete the task without AI. We first compare task composition across individual users, organizational users, and OpenAI workers. We then examine whether users are delegating more complex tasks over time. Finally, we compare task mix across personas, job-title groups, and seniority levels to show how Codex use differs across worker types.

\subsection{Task shares for Codex usage}

To measure what users delegate to Codex, we classify Codex requests into a fixed two-level task taxonomy. The classifier assigns each request to a primary task category based on the user's requested outcome, rather than on incidental tool use during execution. The top-level categories include software-related work, such as code implementation, code understanding, code validation, engineering operations, and application management, as well as broader knowledge-work categories such as data analysis, research, knowledge artifacts, collaboration, and business-function workflows. The full prompt and label definitions are reported in Appendix~\ref{app:task_classifier_prompt}. We use this taxonomy to compare the tasks Individual, Organizational, and OpenAI users delegate to Codex. 

\begin{figure}[!t]
    \centering
    \caption{\raggedright Share of Codex usage by first-task for the average user}
    \label{fig:high-task-shares}

    \includegraphics[width=\linewidth]{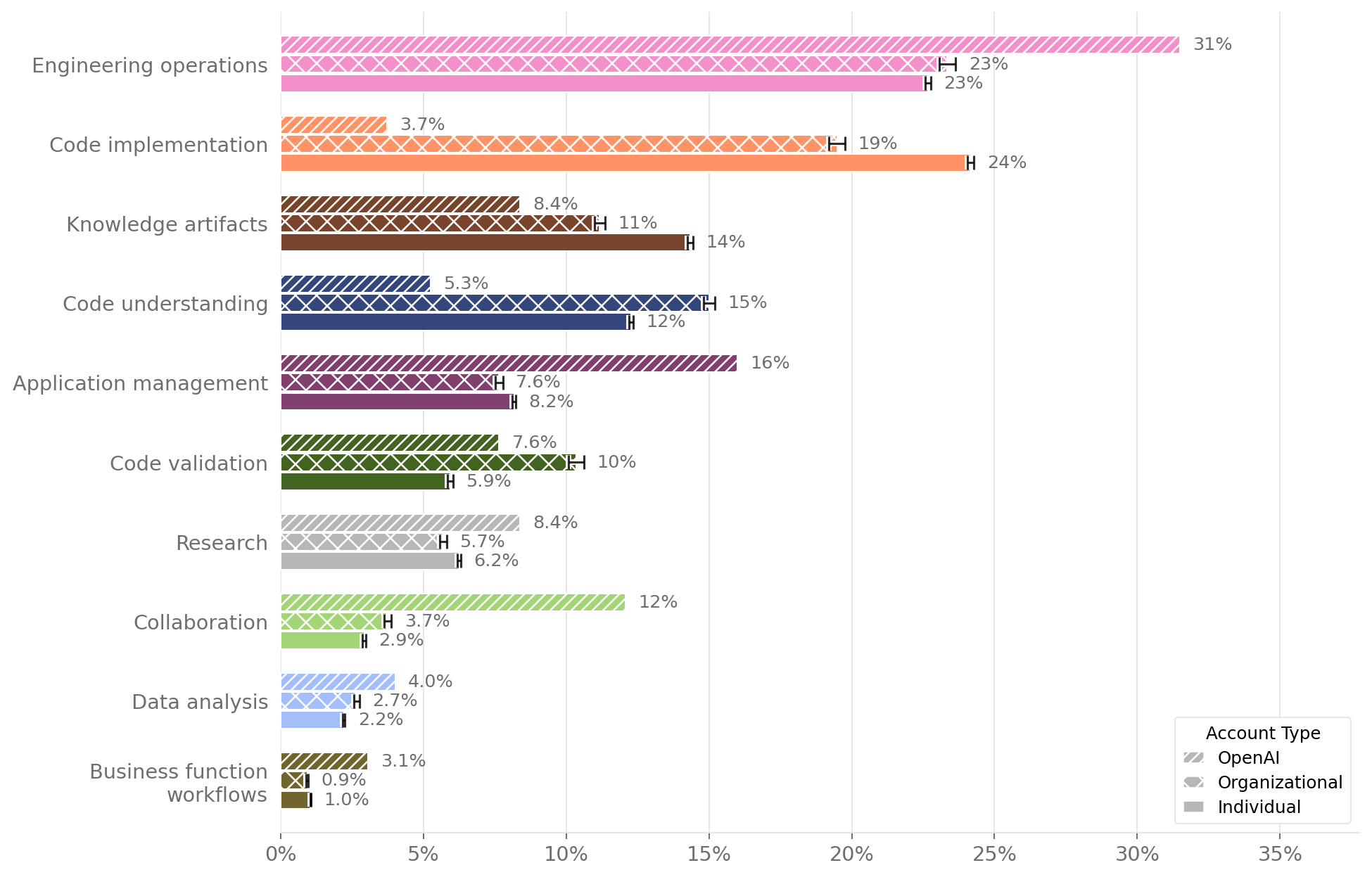}

\begin{minipage}{\linewidth}
\footnotesize{Shares reflect the average user-level share of classified messages falling into each first-level task category during the 28 days prior to \var{codex_analysis_end_date}. The prompt used for our classifications is provided in Appendix~\ref{app:task_classifier_prompt} Messages classified as ``other'' are excluded from the calculation. Organizational and Individual calculations are based on a 4\% sample of users; users inactive during the period are excluded. Error bars show 95\% cluster bootstrap intervals from resampling sampled Individual and Organizational users; intervals reflect sampling uncertainty.}
\end{minipage}

\end{figure}

In Figure~\ref{fig:high-task-shares}, we compare the composition of top-level task usage across account categories, measuring each user's distribution of classified Codex requests and then averaging within account category.\footnote{Appendix~\ref{app:additional-tasks} discusses complementary ways of analyzing task usage, including measuring the share of users who use each category rather than the average user's task mix, and decomposing task use into more granular ``second-level'' categories.} Across all account types, the most common task categories are tied to software production. Engineering operations, code implementation, code understanding, application management, and code validation account for a large share of Codex activity across groups. This indicates that Codex use, while concentrated in software production, is not limited to producing new code. Users also rely on it to understand existing code, validate code changes, manage applications, and support engineering workflows. Importantly, this suggests that Codex is integrated into the broader software development life cycle rather than used only as a code-generation tool.

At the same time, firms closer to the frontier of Codex adoption also use AI for more experimental purposes. Organizational users are more concentrated in coding activities, while OpenAI users show more activity in research, business function workflows, and engineering operations. The pattern suggests that OpenAI use may involve more experimental, research-oriented, or infrastructure-adjacent activity, whereas organizational use is more centered on direct software implementation and understanding.

\subsection{Task complexity on Codex}

The scope of possible requests within each task category is broad: the Code Implementation category covers queries requesting one-line tweaks as well as queries requesting the creation of full-fledged software applications. We quantify the complexity of users' use cases using a prompt, which analyzes the text of a subset of queries and estimates the time that it would take for an experienced human worker to complete the task without the assistance of AI. We use the classifier to assign estimated task durations to the prompts sent by a 0.1\% random sample of Individual accounts who have opted to allow their data to be used for model training. By observing the estimated completion time of turns since November 2025, we are able to measure how the complexity of tasks performed by Codex has changed, as the underlying models have improved and humans have gained experience in assigning tasks to Codex. We provide the prompt used, as well as validation metrics, in Appendix \ref{app:query_complexity_prompt}.

\begin{figure}[!t]
    \centering
    \caption{\raggedright Model-estimated complexity of Codex turns among Individual account users}
    \label{fig:task_complexity}

\begin{subfigure}[t]{0.475\linewidth}
    \centering
    \includegraphics[width=\linewidth]{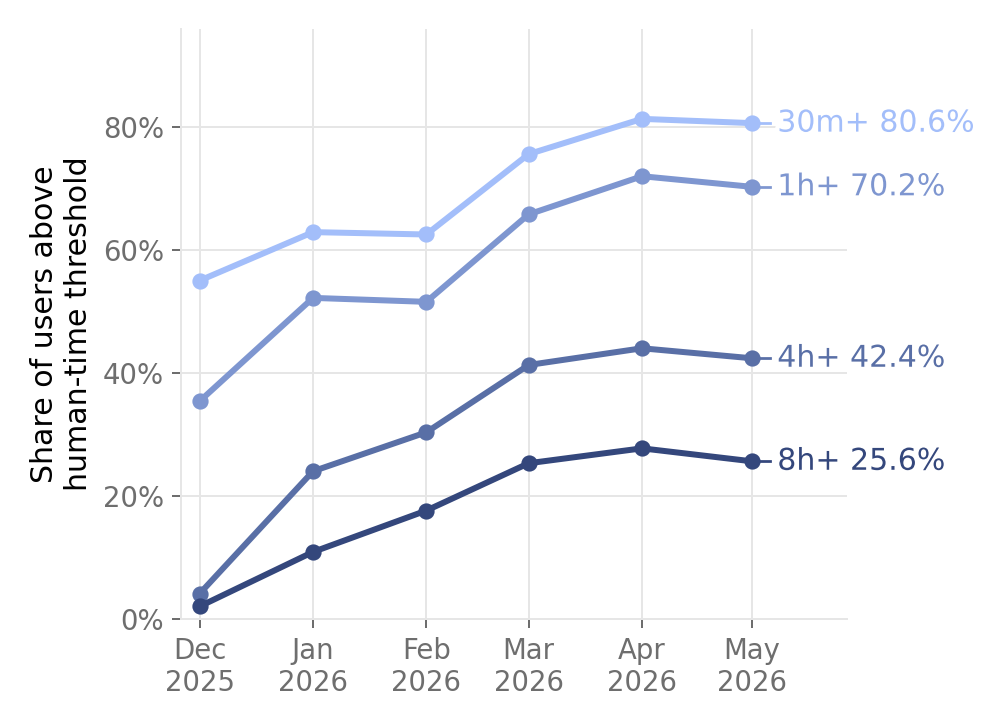}
    \caption{Distribution of users by peak task complexity}
    \label{fig:task_complexity_over_time}
\end{subfigure}
\hspace{0.5cm}
\begin{subfigure}[t]{0.475\linewidth}
    \centering
    \includegraphics[width=\linewidth]{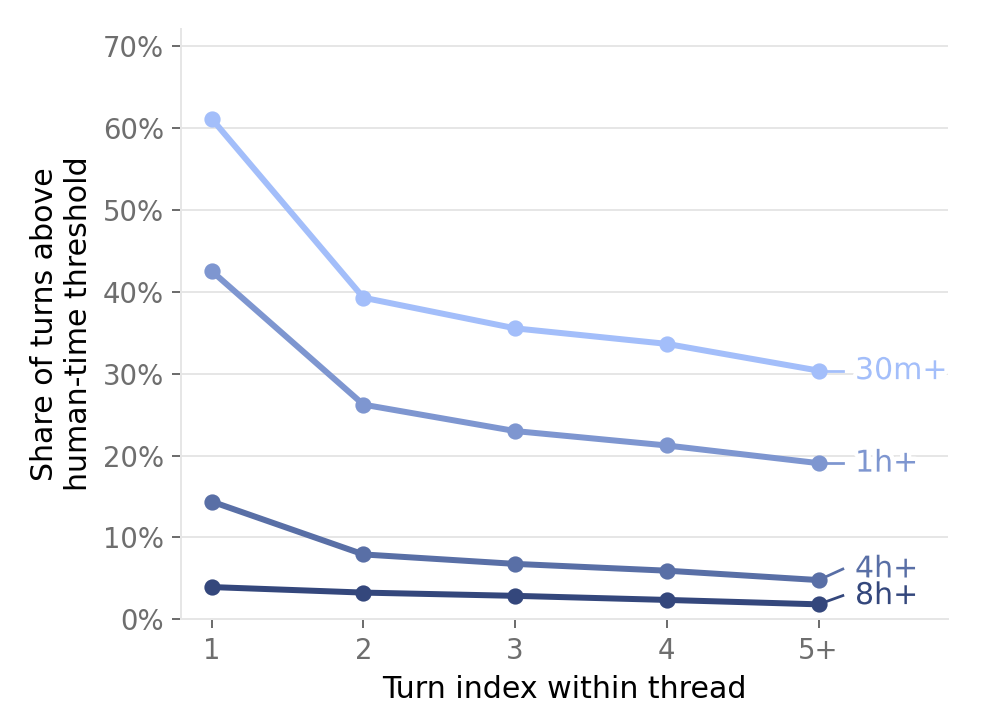}
    \caption{Complexity of turns by turn index}
    \label{fig:task_complexity_by_turn_index}
\end{subfigure}

\begin{minipage}{\linewidth}
\footnotesize{Durations reflect the time it would take an experienced human to complete each task, estimated using the classifier described in Section \ref{app:query_complexity_prompt}. Each panel is constructed using a sample of 0.1\% of Individual users who have opted to allow their queries to be used in training. Figure \ref{fig:task_complexity_over_time} presents the share of sampled users each month that ask at least one query over each threshold of estimated human completion time. The data for February and March is partial. Figure \ref{fig:task_complexity_by_turn_index} shows the share of turns estimated to have a human completion time over certain thresholds, split by the query's position in the thread. Only messages sent in May 2026 by Individual users in the 0.1\% sample are included. An index of 1 corresponds to the first message sent in a thread.}
\end{minipage}

\end{figure}

In Figure \ref{fig:task_complexity}, we present statistics about the distribution of query complexity. Over time, the complexity of users' requests has increased substantially---in December 2025, only \var{consumer_request_complexity_1h_plus_user_share_dec_2025_pct}\% of active Individual users sent at least one prompt that we estimate would have taken an experienced human at least one hour to complete. This share has increased rapidly---in May 2026, we estimate that \var{consumer_request_complexity_1h_plus_user_share_may_2026_pct}\% of users sent at least one prompt of similar or greater difficulty. We estimate even larger proportional increases in the frequency at which users send more difficult prompts: since December 2025, the share of users who send at least one prompt that would have taken an experienced human 8 hours to complete has risen from \var{consumer_request_complexity_8h_plus_user_share_dec_2025_pct}\% to \var{consumer_request_complexity_8h_plus_user_share_may_2026_pct}\%. 

The most complicated queries are concentrated at the start of users' sessions. In Figure \ref{fig:task_complexity_by_turn_index}, we show that the first turn of a thread is more than twice as likely as the fourth turn to require action that would take an experienced human more than one hour. This suggests that users tend to request the broadest, most complex portion of work initially, with further, narrower refinements following later in the session. 

\begin{figure} [!t]
    \centering
    \caption{\raggedright Differences in Codex use cases across different types of workers, both internally at OpenAI (top) and in organizations with job title coverage (bottom)}
    \label{fig:codex_task_share_title_comparison}

    \includegraphics[width=\linewidth]{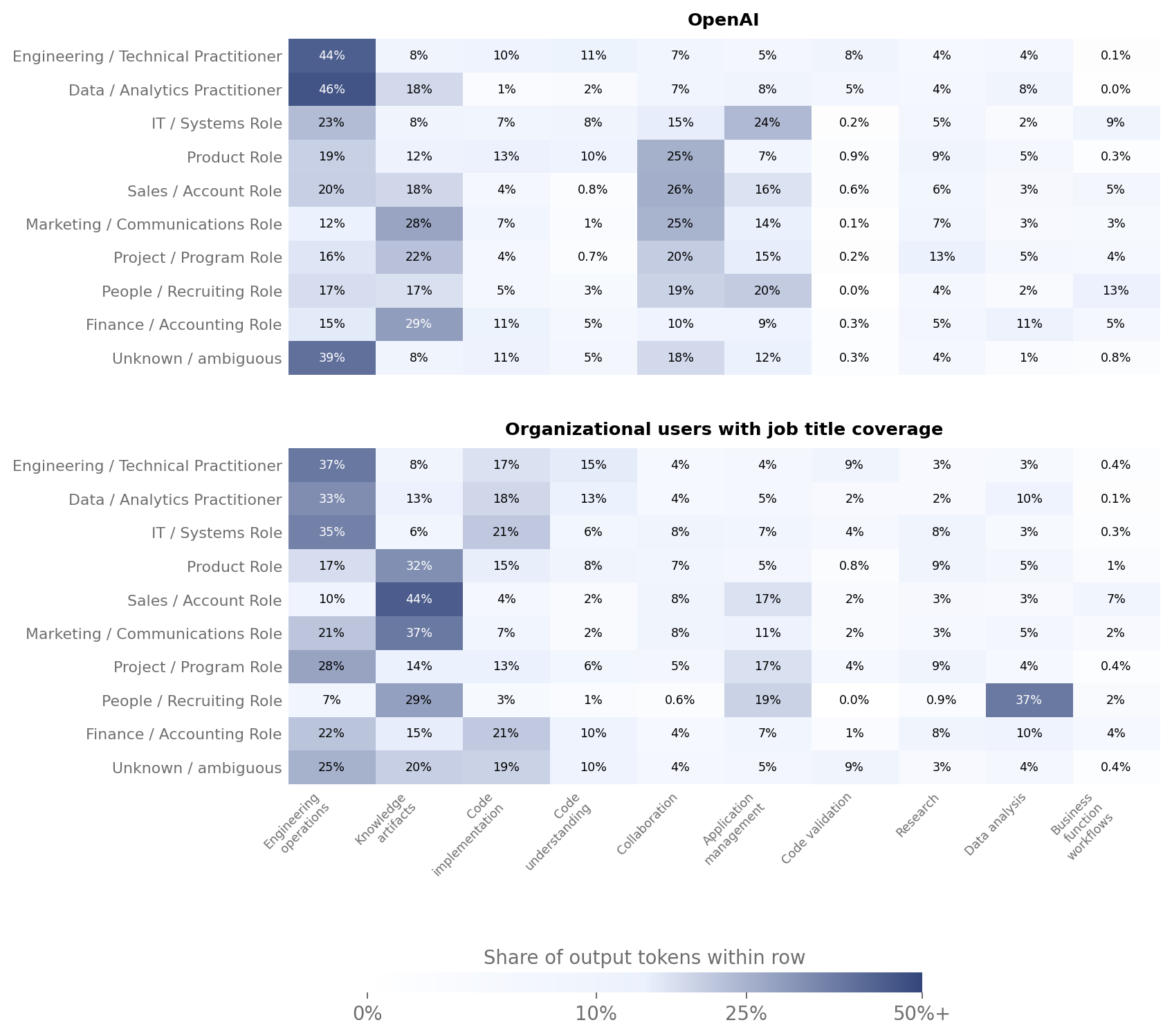}

    \begin{minipage}{\linewidth}
    \footnotesize{Distribution of Codex tokens across task types, by inferred job title, within OpenAI (top) and organizations with job title coverage (bottom).}
    \end{minipage}
\end{figure}

\begin{figure} [!t]
    \centering
    \caption{\raggedright Differences in Codex use cases by seniority level and account type}
    \label{fig:codex_task_share_seniority_comparison}

    \includegraphics[width=\linewidth]{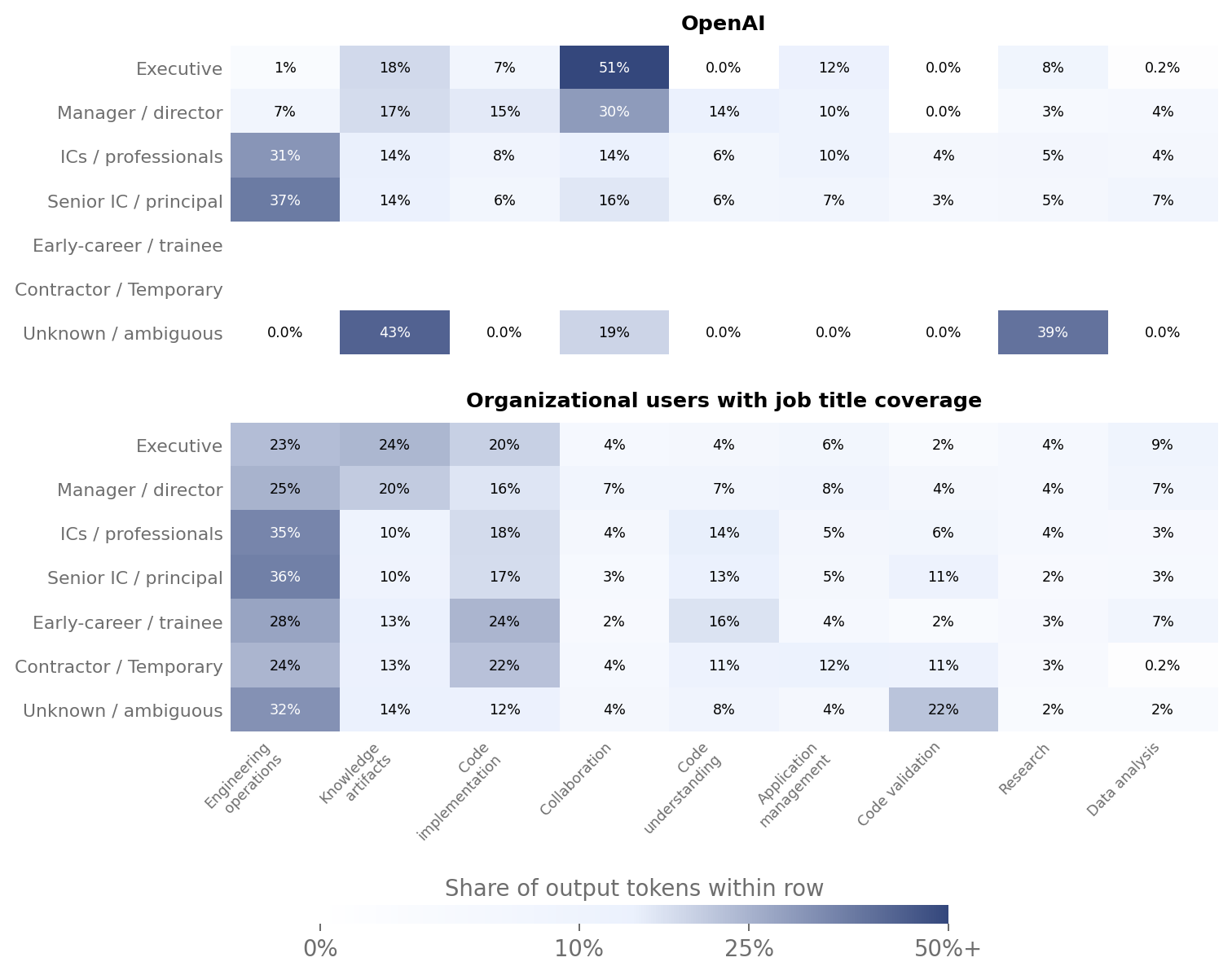}

    \begin{minipage}{\linewidth}
    \footnotesize{Distribution of Codex tokens across task types, by inferred seniority level, within OpenAI (top) and organizations with job title coverage (bottom).}
    \end{minipage}
\end{figure}

\subsection{Task shares by worker type}

We next ask whether the task mix differs across the worker groups studied in Section~\ref{sec:who}. Combining the task classifier introduced above with job title and seniority measures allows us to compare the composition of Codex use within each group. Because the figures report within-group token shares, they describe differences in what users delegate to Codex rather than differences in overall adoption or usage intensity.

Figure~\ref{fig:codex_task_share_title_comparison} compares the composition of Codex use across job title groups, separately for OpenAI workers and users in organizations with job title coverage. We see that most usage both inside OpenAI and inside organizations more broadly is concentrated in engineering operations. Inside OpenAI, across developer and non-developer roles, knowledge artifacts, collaboration, and application management are common tasks. In organizations, many workers are using agentic tools for generating knowledge artifacts, especially in sales, marketing, and recruiting functions. 

Figure~\ref{fig:codex_task_share_seniority_comparison} compares the distribution of Codex output tokens across task categories by inferred worker seniority, separately for OpenAI employees and for workers within organizations with job title coverage. We find that Codex use differs meaningfully across seniority groups, but the patterns do not support a simple interpretation in which junior workers use Codex primarily for code generation while senior workers use it primarily for managerial tasks. In organizations, engineering operations accounts for the largest share of output tokens for most seniority groups, while code implementation is especially prominent among early-career workers and contractors. Knowledge artifacts account for a larger share of tokens among executives and managers, suggesting that more senior workers use Codex partly to produce, interpret, and organize technical documentation and other written outputs. Within OpenAI, the distribution is more uneven: Individual Contributors (ICs) and senior ICs allocate relatively large shares of tokens to engineering operations, while managers and executives devote larger shares to collaboration-oriented tasks. Overall, the figure shows that workers across the seniority distribution use Codex for a broad portfolio of technical and organizational activities, including engineering operations, implementation, code understanding, validation, research, data analysis, and documentation. 

\section{How Are People Using Agentic AI?}
\label{sec:how}

Although agentic AI can already perform tasks within existing production processes, the literature on general-purpose technologies suggests that the largest productivity gains often arise when firms reorganize production around the new technology rather than merely substitute it into existing workflows. We draw here on the canonical example in \textcite{david1990dynamo}, who studies the delayed productivity gains from the transition from steam power to electric power in manufacturing. In the early stages of electrification, many factories replaced centralized steam engines with centralized electric motors while preserving existing factory layouts and work patterns. The larger gains emerged only when firms redesigned production around the distinctive capabilities of electric power: smaller motors made it possible to decentralize power, reorganize the factory floor, change the sequencing of tasks, and adopt more flexible production layouts. These changes took decades to diffuse, partly because firms had to incur substantial fixed costs to redesign plants and develop complementary organizational practices. The history of technological adoption suggests that the near-term effects of agentic AI may understate its longer-run productivity potential if firms have not yet discovered, adopted, or scaled the new production processes that the technology makes feasible.

Building on this literature, we next examine whether users are beginning to adapt their workflows to the capabilities of agentic systems. Having shown what tasks users delegate to Codex, we now study how that delegation is organized. We focus on three observable margins: whether users run multiple agents in parallel, whether agents perform longer blocks of work on users' behalf, and whether users systematize repeated work through skills and plugins. These measures do not fully capture organizational redesign, but they provide early evidence on whether agentic AI use is moving beyond one-off assistance toward more persistent, repeatable, and parallel workflows. The speed of this adjustment may reflect an important difference between digital production processes and earlier episodes of technological change. Whereas electrification often required firms to redesign physical plants and replace durable capital, agentic AI allows workers and organizations to experiment with new workflows at relatively low cost. This lower cost of experimentation may allow new production methods to diffuse more quickly than in prior general-purpose technology transitions, even though the full organizational complements to agentic AI are still emerging.

\subsection{Turn Concurrency}

\begin{figure}[!t]
    \centering
    \caption{\raggedright Codex turn concurrency}
    \label{fig:peak_concurrent_turns}
        \includegraphics[width=.9\linewidth]{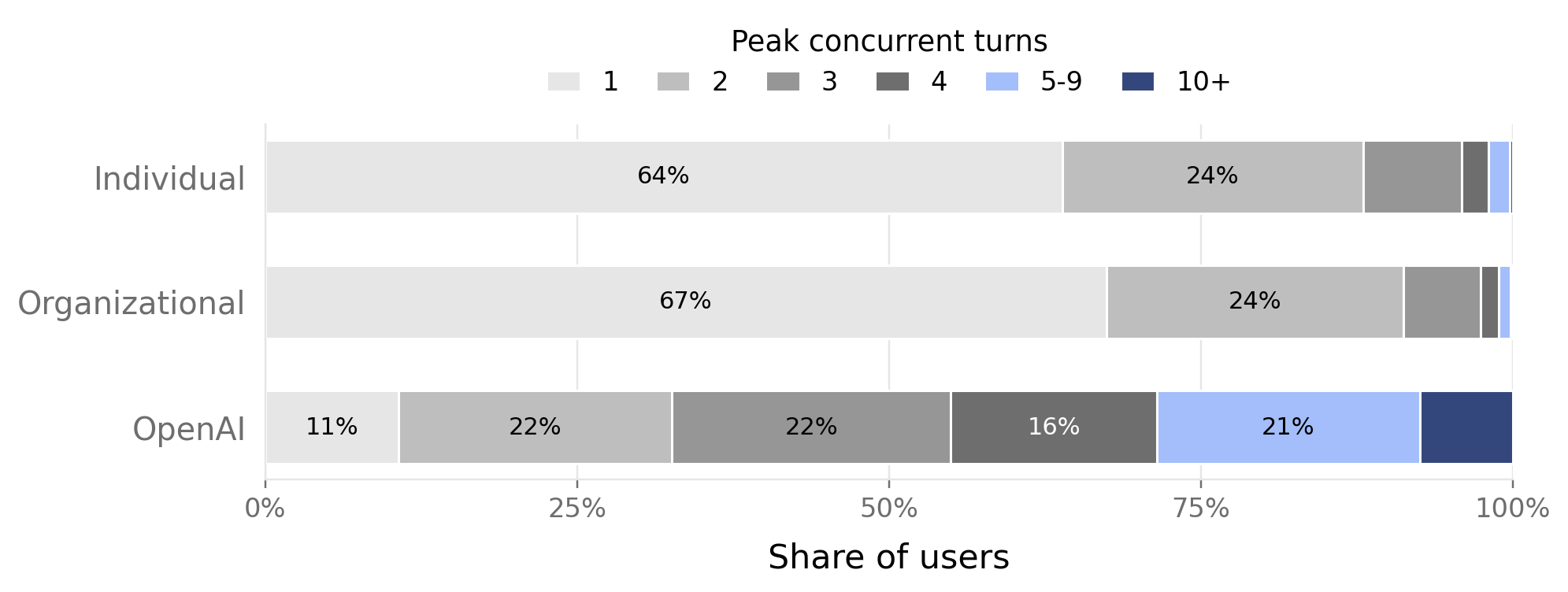}
    \begin{minipage}{\linewidth}
    \footnotesize{Figure summarizes concurrent Codex usage in the week prior to \var{codex_analysis_end_date}, plotting the distribution, across account types, of users' peak number of concurrent agents. Simultaneous agents are measured using overlapping turns, spread across different threads, which overlap for more than 30 seconds}
    \end{minipage}
\end{figure}

Codex, like many AI agents, uses a threaded interaction model in which users can initiate multiple agents and interact with each one in a largely independent workspace. This design changes the structure of delegation. Because long-running tasks may require an agent to work independently for minutes or hours, users need not wait for one task to finish before beginning another. Instead, they can assign several tasks to different agents, monitor progress across threads, and intervene selectively when an agent requires clarification, review, or correction. The threaded model therefore allows users to manage multiple streams of agentic work in parallel, expanding the scope for concurrent task execution relative to conventional single-threaded interactions with AI systems.

It has been speculated that, as agentic tooling advances, parallel workflows will become more common, increasing the value of skills such as delegation and management \parencite{alekseeva2026artificial}. In Figure~\ref{fig:peak_concurrent_turns}, we explore the degree to which users are currently using Codex in parallel fashion. For each user, we calculate the number of overlapping turns they have in different threads during the week prior to \var{codex_analysis_end_date}.\footnote{We restrict only to turns which run in different threads, and which overlap one another for at least 30 seconds.} Each bar shows the distribution of peak concurrency during that week, among users of each account type. Among Organizational and Individual users, the usage of concurrent turns is fairly minimal. Roughly \var{enterprise_turn_concurrency_peak_1_user_share_pct}\% of Organizational users and \var{consumer_turn_concurrency_peak_1_user_share_pct}\% of Individual users do not use concurrent turns at all during this period, and among those who do, the majority peak at two concurrent turns.  Among OpenAI users, however, workflows have become far more parallelized. Only \var{openai_turn_concurrency_peak_1_user_share_pct}\% of users in this group use  a sole workflow at any one time during this period, and nearly \var{openai_turn_concurrency_peak_5plus_user_share_pct}\% managed five or more concurrent agents at some point during this period. The prevalence of this practice within OpenAI represents a workflow in which humans oversee a team of agents, delegating tasks across many simultaneous workers at once. This workflow is fundamentally different from the one practiced by most external users, in which humans directly perform the majority of work, as it requires workers to manage, delegate to, and review the work of a relatively large group of agents.

\subsection{Long-running agents}

\begin{figure}[htb]
    \centering
    \caption{\raggedright Cumulative Codex turn duration per day}
\begin{subfigure}[t]{0.475\linewidth}
    \centering
    \includegraphics[width=\linewidth]{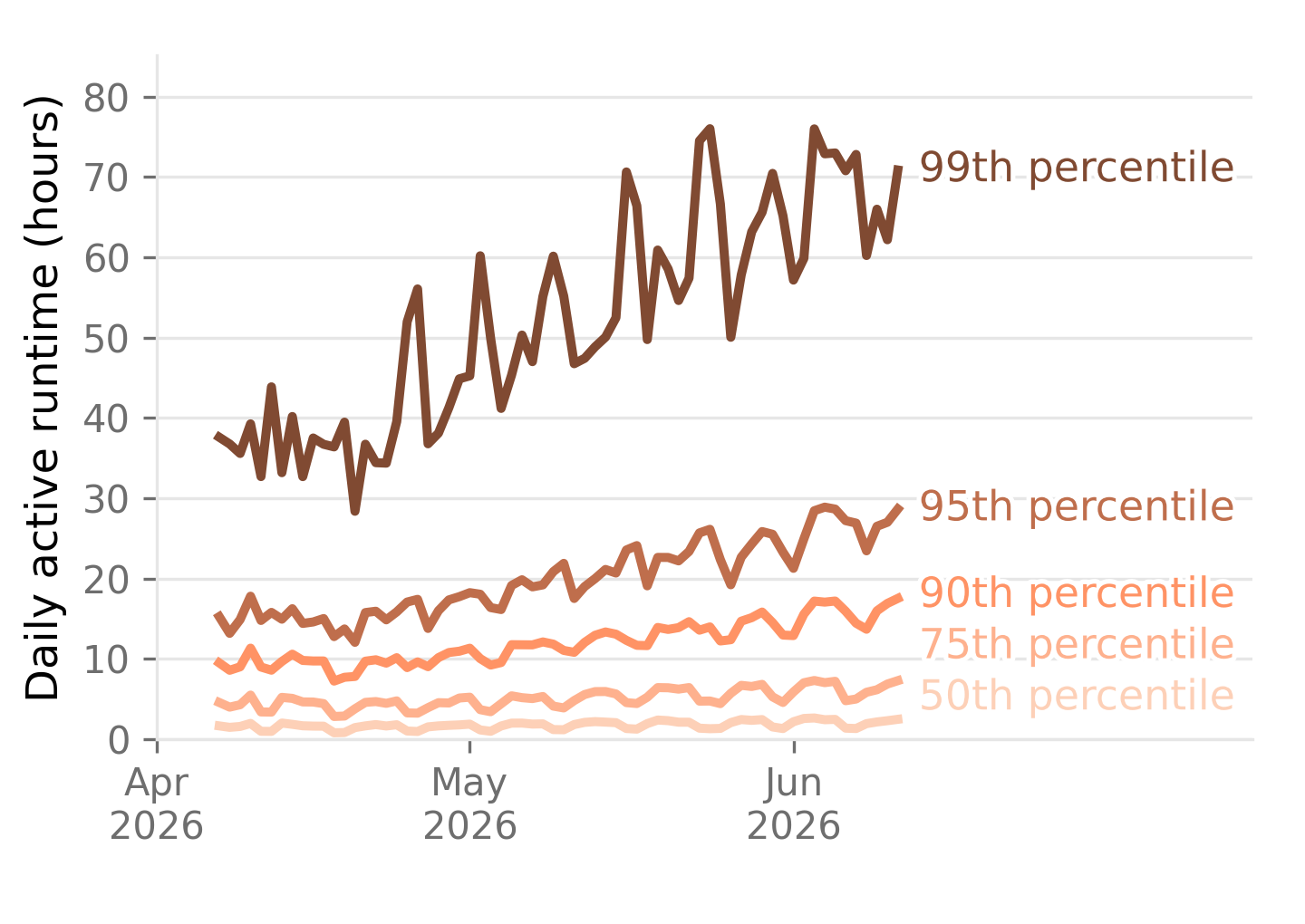}
    \caption{OpenAI users}
    \label{fig:sum_runtime_openai}
\end{subfigure}
\hspace{0.5cm}
\begin{subfigure}[t]{0.475\linewidth}
    \centering
    \includegraphics[width=\linewidth]{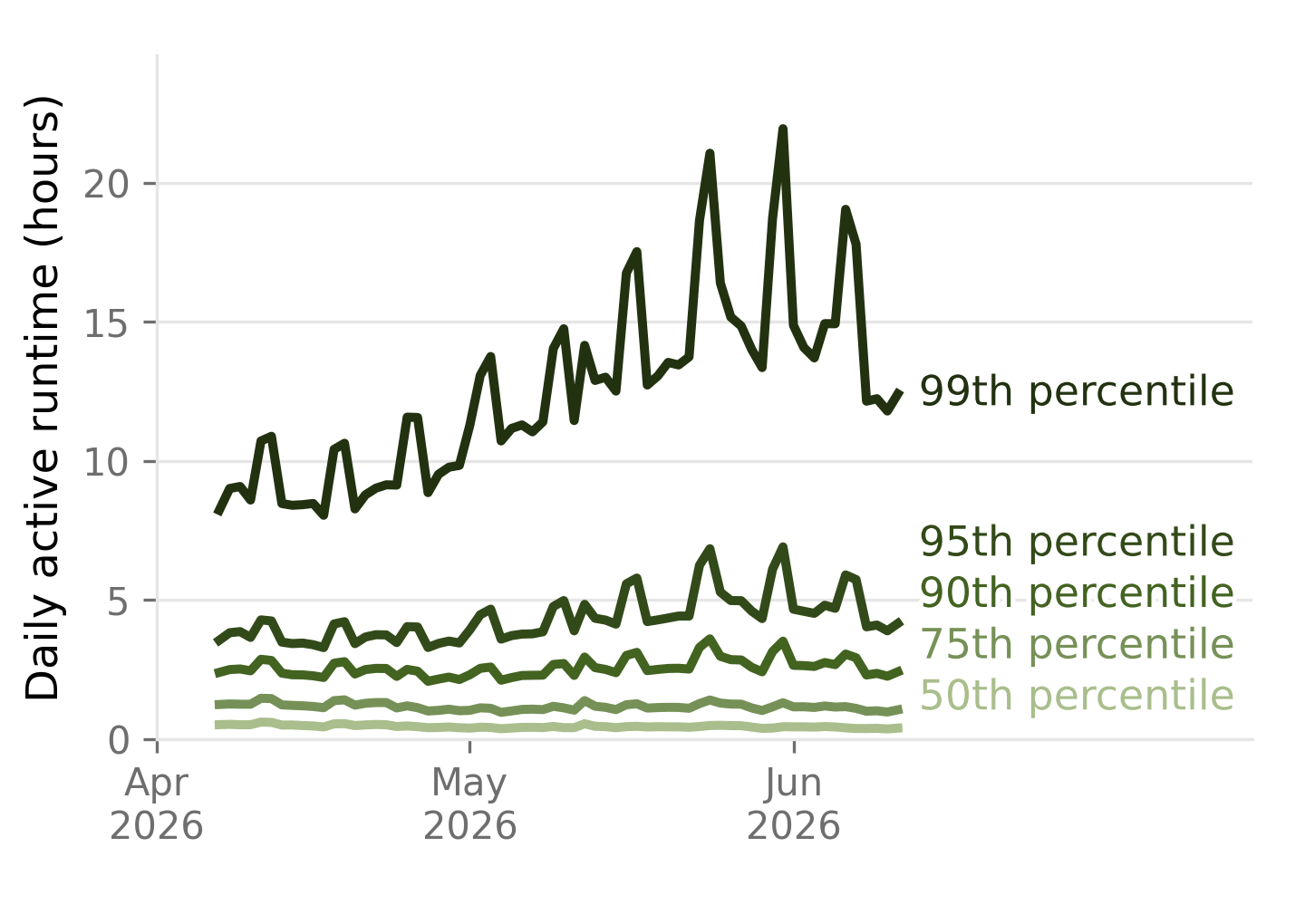}
    \caption{Organizational users}
    \label{fig:sum_runtime_enterprise}
\end{subfigure}

\begin{subfigure}[t]{0.475\linewidth}
    \centering
    \includegraphics[width=\linewidth]{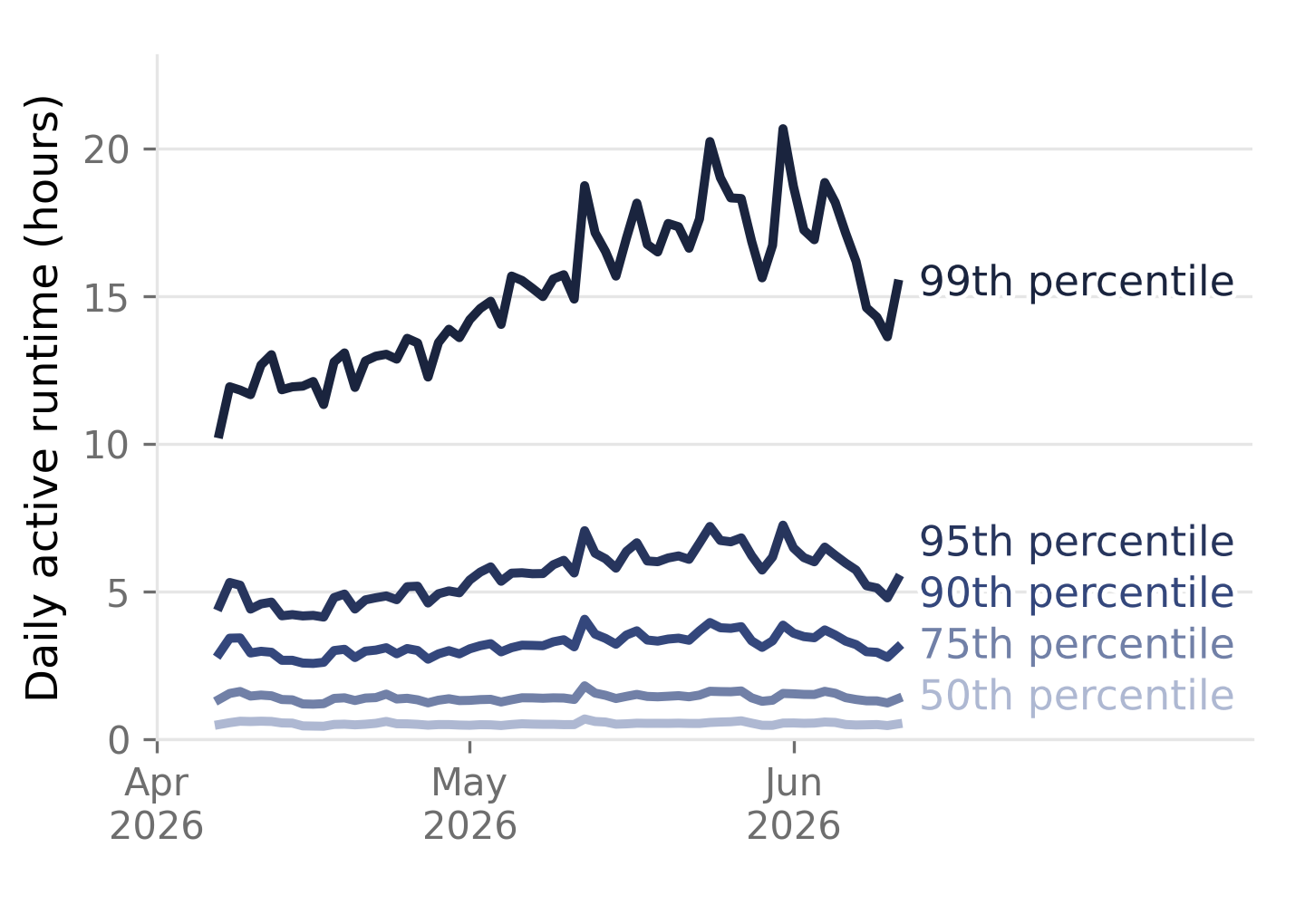}
    \caption{Individual users}
    \label{fig:sum_runtime_consumer}
\end{subfigure}
\begin{minipage}{\linewidth}
    \footnotesize{Figure summarizes total active runtime of users' Codex turns on each day. This measure excludes gaps where no output is returned for more than 30 minutes, which predominantly reflect periods during which Codex is awaiting user input. Because overlapping turns are summed, a user's cumulative runtime during a day can exceed 24 hours. Figure \ref{fig:sum_runtime_openai} plots usage among OpenAI accounts, Figure \ref{fig:sum_runtime_enterprise} plots usage among Organizational accounts. Figure \ref{fig:sum_runtime_consumer} plots usage among Individual accounts. Accounts inactive on each day are excluded.}
    \end{minipage}
\end{figure}

Parallelism captures one way agentic use departs from conversational interaction; another is the duration of delegated work. Researchers and practitioners have emphasized the possibility that agents can operate asynchronously for extended periods with limited human supervision. 

In Figure \ref{fig:sum_runtime_openai}, we examine whether this possibility has begun to reshape workflows within OpenAI by measuring the amount of time Codex tasks remain active over the course of a day. For the median OpenAI employee, agentic workflows remain intermittent: on \var{codex_analysis_end_date}, the median employee had Codex turns running for \var{openai_daily_task_runtime_p50_hours_end_date} hours. The typical user therefore delegates meaningful blocks of work, but has not shifted to continuous, around-the-clock autonomous execution.\footnote{We measure the time an agent is active within each turn by summing the latency of all completed request-response pairs. We include gaps of up to 30 minutes between responses, which include the results of tool calls, reasoning, and the output text the agent returns. We remove longer gaps, which are often driven by agents who are awaiting user input or paused. Each user is assigned the total duration of all agent turns they initiate and which begin within that calendar day. Turns that are not interactive, such as those that are run as part of a batch script, are excluded.} Among frontier users, however, this pattern looks considerably different; OpenAI employees at the 99th percentile of the distribution have recently run about \var{openai_daily_task_runtime_p99_hours_end_date} hours of agent turns within the average day, which implies that at a given hour of the day, they have several agents running concurrently. These parallel, long-running workflows have increased in prevalence recently, with 99th percentile users within OpenAI increasing their cumulative daily runtime by nearly \var{openai_daily_task_runtime_p99_growth_since_start_pct}\% since \var{task_runtime_daily_series_start_date}. During the period we observe, the absolute increase in daily runtimes within OpenAI was largest at the top of the distribution, although lower percentiles grew faster proportionally.

Outside OpenAI, cumulative daily runtime remains substantially lower, and the median user continues to use Codex only intermittently over the course of the day. We nevertheless observe substantial growth at the upper tail of the usage distribution during our sample period. At the 99th percentile, daily runtime increased by roughly 25 percent among Organizational users and by about 50 percent among Individual users. While usage intensity is rising across all user populations, these patterns show that OpenAI employees use Codex much more intensively than external users. The contrast between the median and the upper tail suggests that agentic workflows remain sporadic for typical users, but that a smaller group of high-intensity users is rapidly expanding the amount of work it delegates to Codex.

\subsection{Systematization of Agentic Work}

A final margin is whether delegated workflows remain one-off interactions or become reusable routines. The simplest use of agentic AI is \emph{ad hoc}: a user describes a task, the agent performs it, and the interaction ends. Codex, like other agentic AI software, also allows users to codify workflows and procedural preferences so that similar work can be delegated repeatedly. We refer to this shift from \emph{ad hoc} use to reusable workflow infrastructure as the ``systematization'' of agentic work. Systematization is an important step toward delegated production: without it, users must repeatedly supply task context, procedural guidance, and instructions, limiting the extent to which work can be handed off.

\begin{figure}[H]
    \centering
    \caption{\raggedright Skill use over time and by account category}
    \label{fig:skill_user_share_combined}

    \begin{subfigure}[t]{0.9\linewidth}
        \centering
        \includegraphics[width=\linewidth]{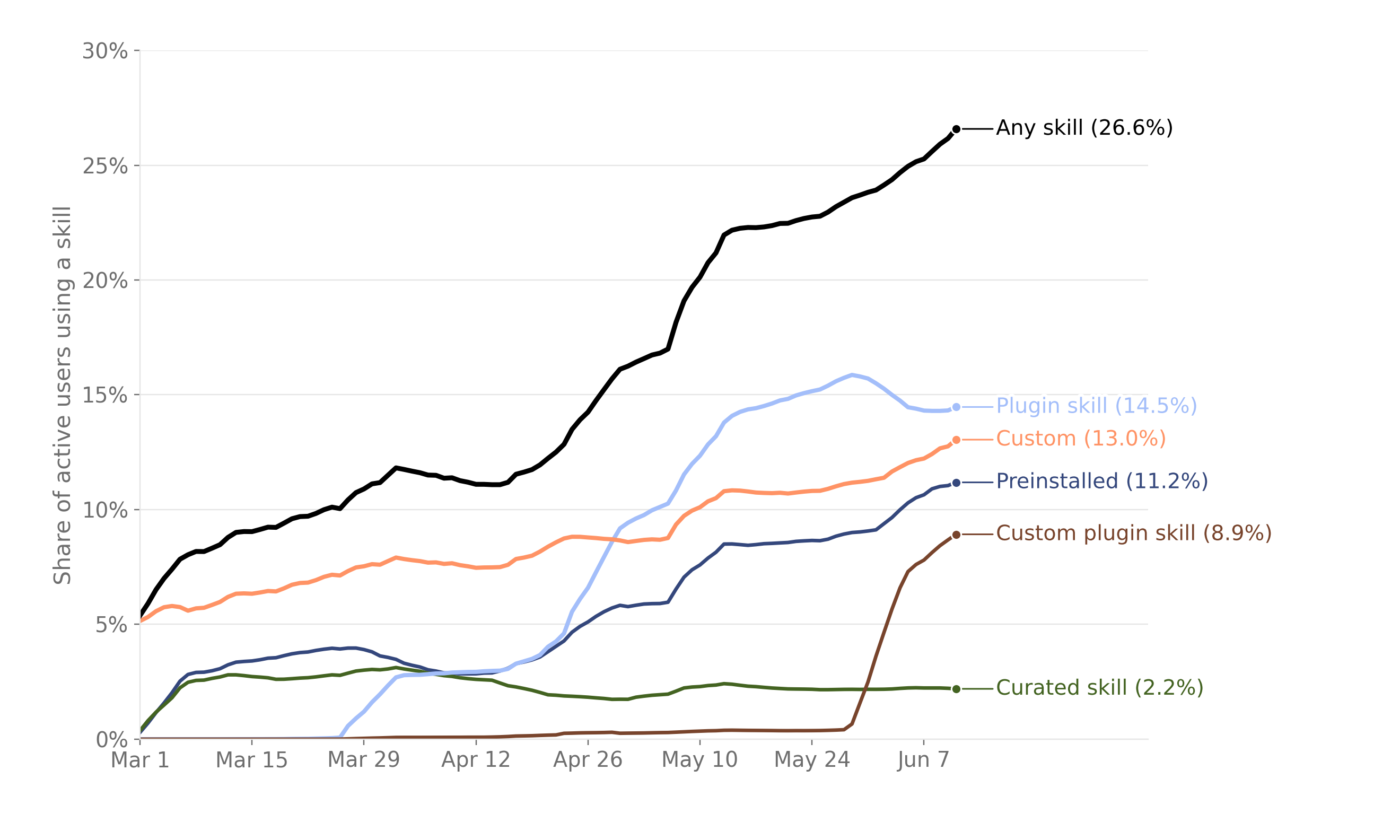}
        \caption{\raggedright Skill use over time}
        \label{fig:skill_user_share_time}
    \end{subfigure}

    \vspace{0.5cm}

    \begin{subfigure}[t]{0.9\linewidth}
        \centering
        \includegraphics[width=\linewidth]{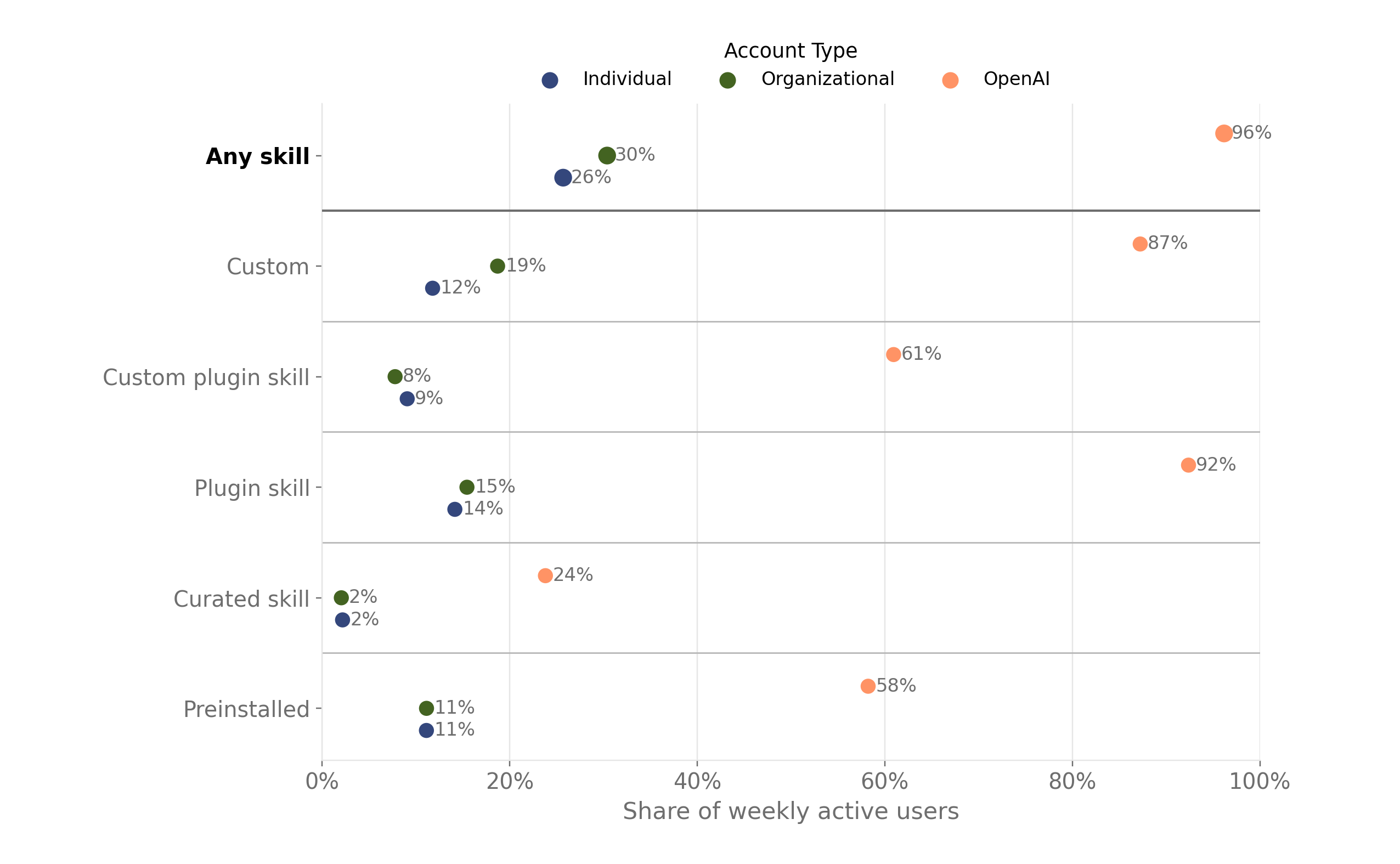}
        \caption{\raggedright Skill use by account category}
        \label{fig:skill_user_share}
    \end{subfigure}

    \begin{minipage}{0.9\linewidth}
    \footnotesize{(a) Each series shows the share of weekly active Codex users who invoked each skill source at least once during the prior \var{skill_usage_window_days}-day window. Users can appear in multiple skill-source categories. (b) Each dot shows the share of weekly active Codex users in an account category who invoked each skill source at least once during the \var{skill_usage_window_days}-day window ending on \var{codex_analysis_end_date}. Users can appear in multiple categories.}
    \end{minipage}
\end{figure}

Within Codex, work is often systematized through skills and plugins, which combine instructions, software, and external-tool integrations that can be reused during a task.\footnote{In OpenAI's terminology, a \emph{skill} is the authoring format for a reusable, task-specific workflow: a directory containing a required \texttt{SKILL.md} file, with \texttt{name} and \texttt{description} metadata and optional scripts, references, and assets. A \emph{plugin} is an installable distribution unit, identified by a \texttt{.codex-plugin/plugin.json} manifest, that can package skills together with app integrations, Model Context Protocol (MCP) configuration, hooks, and supporting assets. Thus, a skill specifies a workflow, whereas a plugin packages and distributes capabilities.} These skills and plugins can be shared across users, allowing workflows to be systematized not only by individuals but also within and across organizations. They therefore make it easier to transfer procedural context, tool access, and task-specific guidance across repeated uses of Codex.

In our analysis, we distinguish between five sources of skills:

\begin{itemize}
\item \emph{Preinstalled} skills, such as image generation, are reusable capabilities bundled with Codex.
\item \emph{Curated} skills, such as PDF handling, are standalone OpenAI-distributed skills that are not tied to a plugin.
\item \emph{Plugin} skills, such as Google Drive document workflows, are bundled into a plugin alongside other skills or software.
\item \emph{Custom plugin} skills are skill invocations that are associated with a recognized plugin but do not match a skill in the curated plugin skill catalog.
\item \emph{Custom} skills are standalone skills not distributed by OpenAI, such as team-specific data visualization guidelines or a research workflow.
\end{itemize}
These categories capture different forms of systematization, ranging from product-provided reusable capabilities to user- or organization-specific workflow codification.

Figure~\ref{fig:skill_user_share_combined} shows that skill use is already common, increasing over time, and unevenly distributed across account categories. In the \var{skill_usage_window_days}-day window ending on \var{codex_analysis_end_date}, \var{consumer_skill_any_user_share_pct}\% of active Individual Codex users and \var{enterprise_skill_any_user_share_pct}\% of active Organizational Codex users invoked at least one skill. Within OpenAI, skill use is nearly universal: \var{openai_skill_any_user_share_pct}\% of active Codex users invoked at least one skill. Skill use also increased substantially during the period shown in Figure~\ref{fig:skill_user_share_time}: the share of active Codex users invoking any skill rose from \var{skill_any_user_share_start_pct}\% on March 1, 2026 to \var{skill_any_user_share_latest_pct}\% on \var{codex_analysis_end_date}.

The growth in skill use comes especially from plugins and custom skills. These two categories represent different forms of systematization. Plugins extend Codex's general-purpose capabilities into recurring task domains such as documents, spreadsheets, slide decks, and other structured artifacts. Custom skills serve a different role: they allow users and organizations to encode more local procedural context, such as team writing standards, recurring reports, organization-specific workflows, or user-specific preferences. The growing use of custom skills suggests that users value not only the model's general capabilities, but also the ability to attach persistent procedural context to repeated tasks that are too specialized to be fully covered by standardized plugins.

Skill use differs across account types not only in terms of overall frequency, but also in terms of the composition of skills invoked. Figure~\ref{fig:skill_user_share} shows that plugin and custom skill usage is highest among OpenAI users and substantially higher among Organizational users than among Individual users. These differences are consistent with variation in the returns to workflow systematization across settings. Custom skills are especially valuable when repeated tasks depend on organization-specific context, shared conventions, internal procedures, or team-level standards. These conditions are more common in Organizational settings than in Individual use. The large differences in custom skill adoption therefore suggest that users derive the greatest value from systematized workflows in high-context organizational environments, where persistent procedural context can reduce coordination costs and improve consistency across repeated tasks.\footnote{Appendix Figure~\ref{fig:skill_use_by_task_area} shows that skill invocation rates also vary by task area. Among Individual and Organizational users, skill use remains concentrated in technical workflows, including application management, collaboration, code validation, and code understanding. Within OpenAI, skill use has diffused further into broader knowledge-work workflows; for example, \var{openai_skill_task_area_collaboration_conversation_share_pct}\% of collaboration conversations invoke at least one skill.}

\section{Conclusion}

We document the emergence of agentic AI as a distinct workplace technology. Much prior research has studied AI as an information, advice, and content-generation tool, where the primary interaction is conversational. Agentic systems introduce a different mode of use. Rather than simply asking for information, users increasingly delegate work: producing artifacts, modifying systems, and executing workflows. This shift suggests that standard measures of AI use, such as active users, chats, or message volume, may become less informative as agentic systems diffuse. Future analyses may need to track delegated task complexity, runtime, workflow reuse, concurrency, and production output.

The evidence we report suggests that the diffusion of agentic AI is not just a shift in which AI tool people use, but a shift in how they organize work around AI. Early use often resembles familiar conversational assistance: asking questions, drafting text, or generating code in a back-and-forth exchange. More agentic use involves assigning tasks, reviewing outputs, coordinating multiple threads of work, and reusing codified workflows. Intensive users appear to manage portfolios of agentic work, with their own role shifting toward delegation, supervision, and integration. This suggests that the adoption of agentic AI involves the development of new work practices, not simply more frequent use of AI.

Software development is the leading edge of this change. It is still the most common use case for agentic AI tools across user populations, likely because software work has properties that lend themselves to agentic AI: software work is digital, produces verifiable artifacts, and contains many modular subtasks. At the same time, observed uses extend beyond code generation. Users are also delegating system understanding, debugging, validation, documentation, configuration, and operational work.

\begin{figure}[h]
    \centering
    \caption{\raggedright Change in median output tokens per person, by OpenAI job function}
    \label{fig:oai_output_token_change}
    \includegraphics[width=\linewidth]{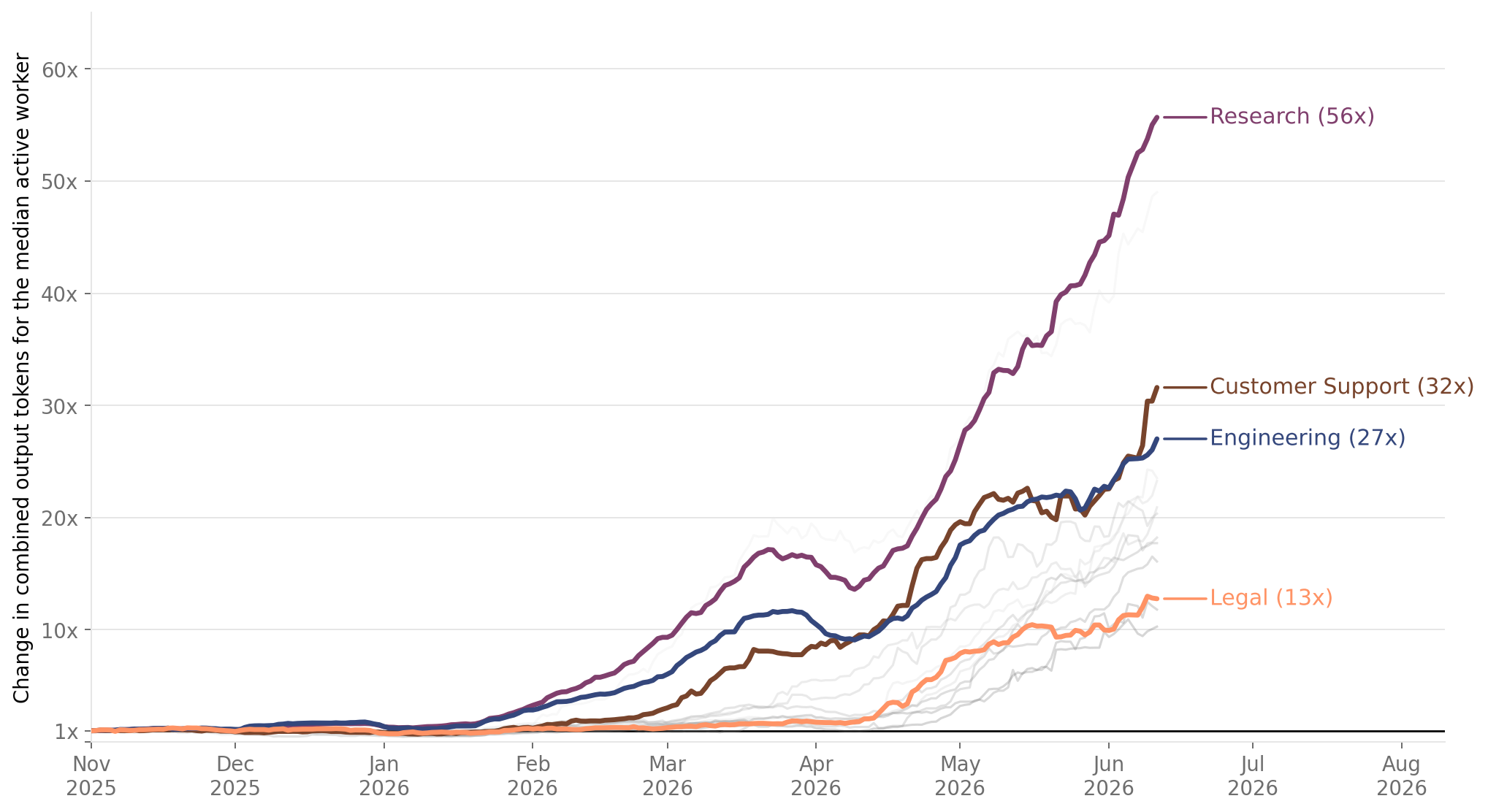}
\begin{minipage}{\linewidth}
\footnotesize{Plot shows the normalized number of output tokens generated in the trailing 28 days by the median worker in each job function within OpenAI. Output tokens include those generated on ChatGPT and those generated on Codex. Each point includes only users active in the preceding 28 days. All series are normalized to have a value of 1 on November 1, 2025.}
\end{minipage}
\end{figure}

We find that as adoption deepens, the scope of agentic work expands. The broadest task portfolios are observed where frictions to adoption are lowest. Among OpenAI workers, Codex is used not only for software engineering but also for research, planning, communication, recruiting, sales, product work, and data analysis. These patterns suggest that while software may be the initial deployment domain for agentic AI, it is diffusing quickly across this boundary.
Figure~\ref{fig:oai_output_token_change} shows that as usage patterns within OpenAI have intensified, so have measures of output---between \var{internal_job_function_relative_token_usage_baseline_date} and \var{codex_analysis_end_date}, the median worker's number of output tokens rose at least \var{internal_job_function_min_p50_relative_l28d_combined_output_tokens_fold}-fold in every job function. This growth in token output reflects several factors, including the shift from conversational interfaces to agentic tools and the later emergence of more intensive, persistent uses of those tools. Our evidence from OpenAI is not necessarily representative of the typical organization, but it shows what agentic AI use can look like when adoption frictions are low: use expands across functions, output rises sharply, and work increasingly takes the form of delegated tasks rather than isolated exchanges.

These patterns have implications beyond adoption and productivity measurement. As AI use shifts from consultation to delegation, organizations may need to rethink how work is allocated, reviewed, and coordinated. Agentic systems make it possible to offload larger, more repeatable, and more modular units of work, potentially reducing the importance of some routine execution tasks while increasing the importance of judgment, oversight, coordination, and review. Jobs may increasingly involve directing, monitoring, and integrating the outputs of AI agents rather than executing each component task directly. Organizations may also redesign workflows around parallelized agent labor, allowing individual workers to manage multiple workstreams at once. These changes could affect team composition, hiring needs, career ladders, and the distribution of work across skill levels.

Taken together, the evidence suggests that agentic AI is not simply a more capable form of conversational AI. Conversational systems are mainly used to exchange information with users; agentic systems are increasingly used to carry out work on users' behalf. The most intensive users do not just chat more often. They delegate tasks, run work in parallel, reuse codified workflows, and shift their own effort toward supervision and integration. The frontier of AI adoption is therefore moving from asking systems for answers toward managing systems that act. Documenting that transition is essential for understanding how AI may reshape work, organizations, and the structure of knowledge production.

\FloatBarrier
\printbibliography

\pagebreak

\appendix

\renewcommand{\thefigure}{A\arabic{figure}}
\renewcommand{\thetable}{A\arabic{table}}
\setcounter{figure}{0}
\setcounter{table}{0}

\section{Additional Figures}

\begin{figure}[H]
    \centering

\caption{\raggedright Codex usage share by people manager status}
    \centering
        \includegraphics[width=\linewidth]{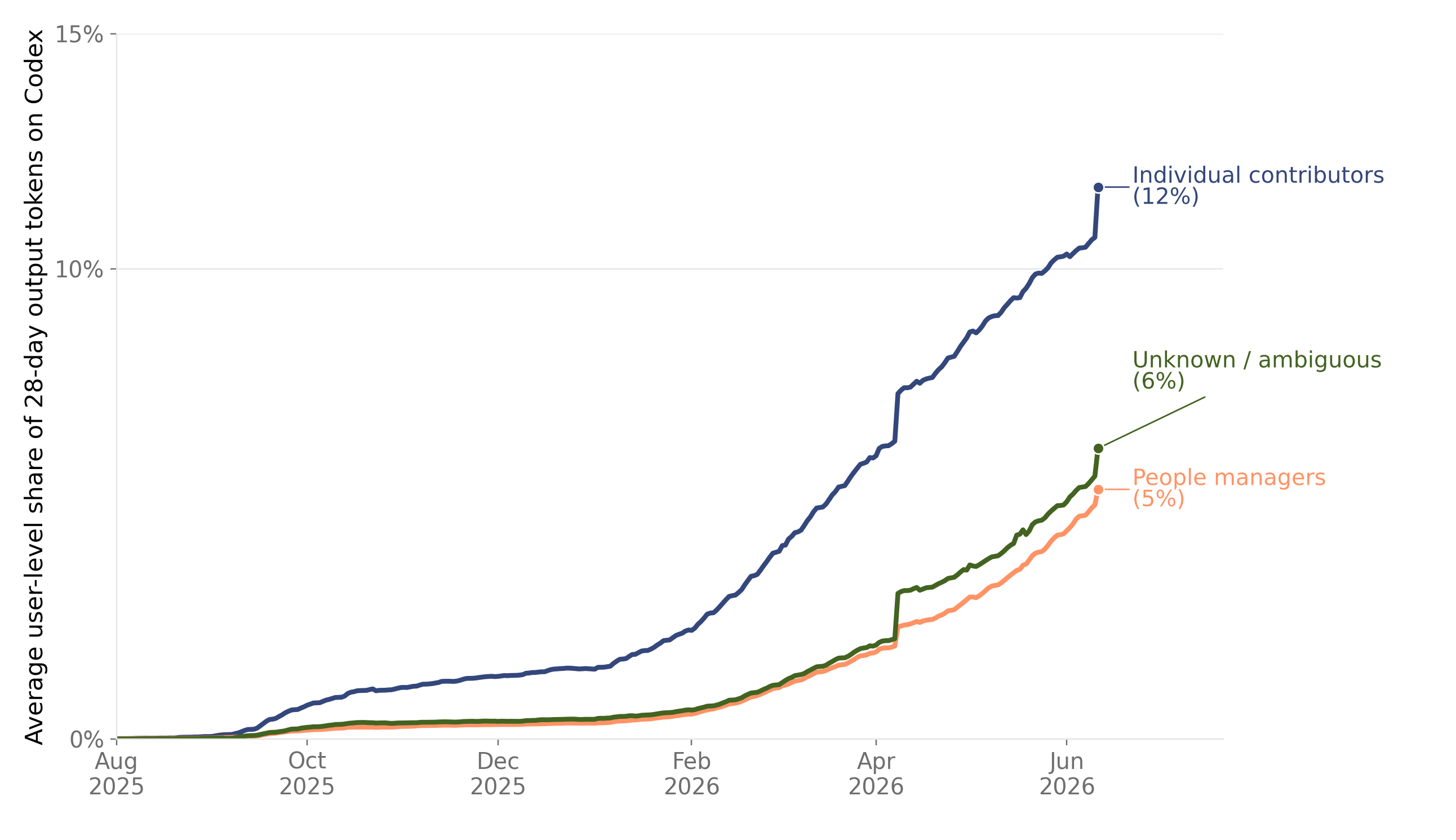}
    \label{fig:codex_token_manager}
    \begin{minipage}{\linewidth}
        \footnotesize{Each series shows the average user-level share of output tokens across ChatGPT and Codex that were produced on Codex for organizational users with high-quality job-title information, by inferred people manager status. Figures are based on output tokens generated on each platform in the preceding 28 days. Only users active in the preceding 28 days are included.}
    \end{minipage}
\end{figure}

\begin{figure}[H]
    \centering
    \caption{\raggedright Users using Codex first-level tasks by account category}
    \label{fig:task_category_user_adoption_shares}
    \includegraphics[width=\linewidth]{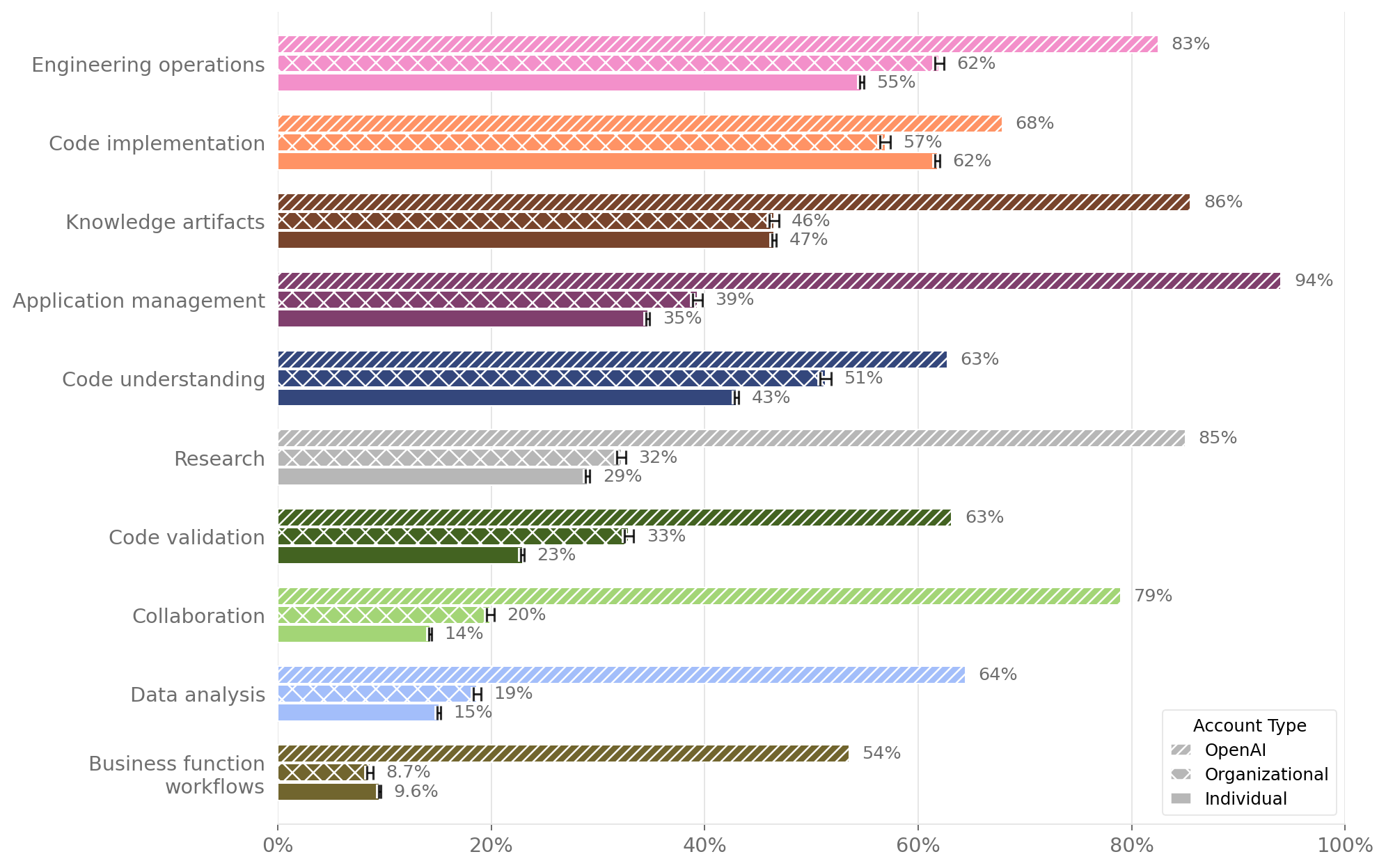}

    \begin{minipage}{\linewidth}
    \footnotesize{Shares reflect the share of users with at least one classified turn in each first-level task category during the 28 days prior to \var{codex_analysis_end_date}. Denominator is users with at least one classified turn during the period. Turns classified as “other” are excluded from the calculation. Calculations are based on a 4\% sample of users. Account categories are based on OpenAI internal-user flags and explicit request plan-type whitelists, with OpenAI taking precedence. Other plan types are omitted from plots. Error bars show 95\% cluster bootstrap intervals from resampling sampled users; intervals reflect sampling uncertainty.}
    \end{minipage}
\end{figure}

\begin{figure}[H]
    \centering
    \caption{\raggedright Users using Codex second-level tasks by account category}
    \includegraphics[width=0.85\linewidth]{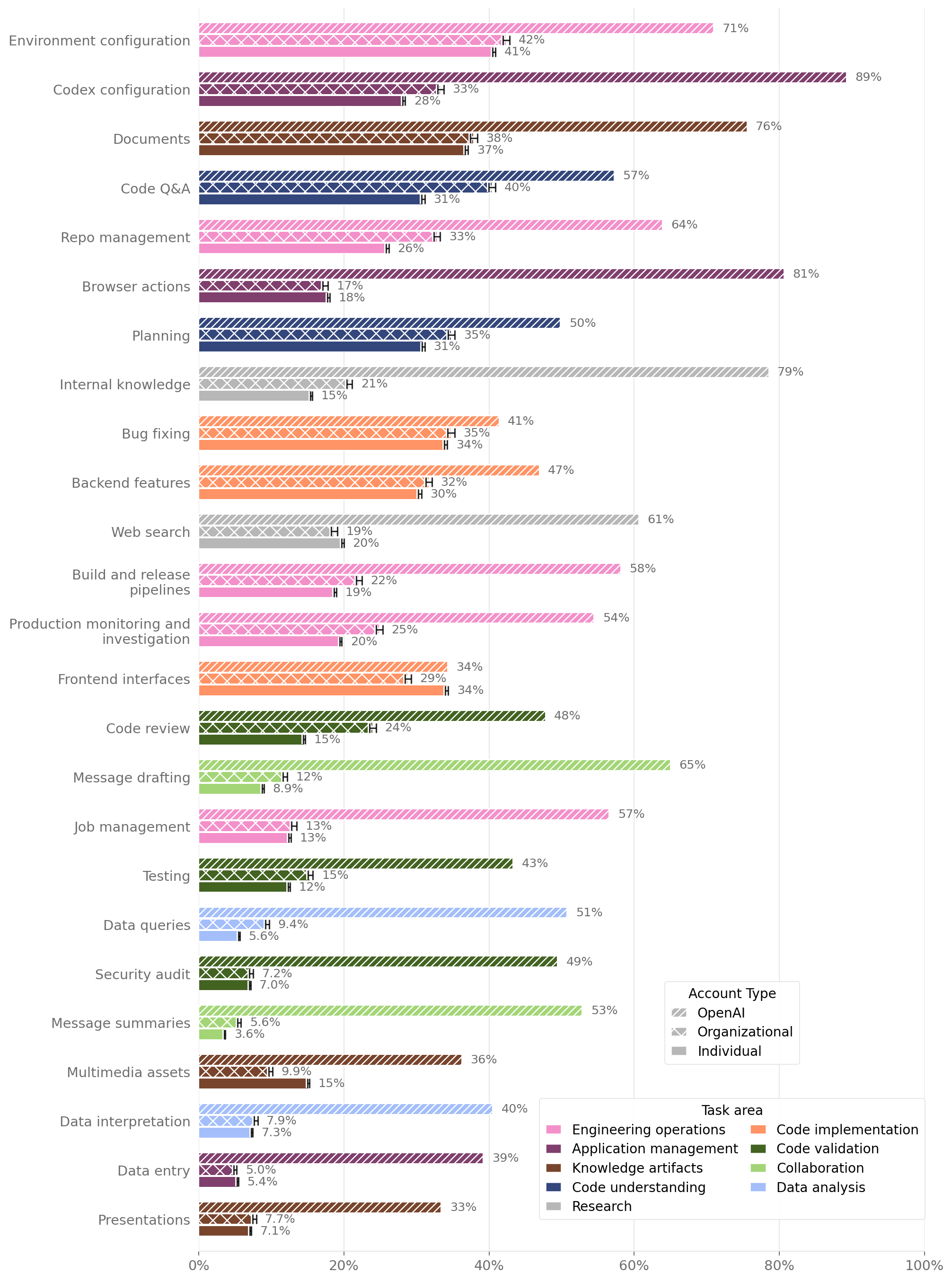}
    \label{fig:mid-task-shares-user-adoption-category-share}

    \begin{minipage}{\linewidth}
    \footnotesize{Shares reflect the share of users with at least one classified turn in each second-level task category during the 28 days prior to \var{codex_analysis_end_date}. Denominator is users with at least one classified turn during the period. Turns classified as “other” are excluded from the calculation. Calculations are based on a 4\% sample of users. Account categories are based on OpenAI internal-user flags and explicit request plan-type whitelists, with OpenAI taking precedence. Other plan types are omitted from plots. Error bars show 95\% cluster bootstrap intervals from resampling sampled users; intervals reflect sampling uncertainty. Top 25 second-level task categories shown.}
    \end{minipage}
\end{figure}

\begin{figure}[H]
    \centering
    \caption{\raggedright Average user second-level task category mix on Codex by account category}
    \includegraphics[width=0.85\linewidth]{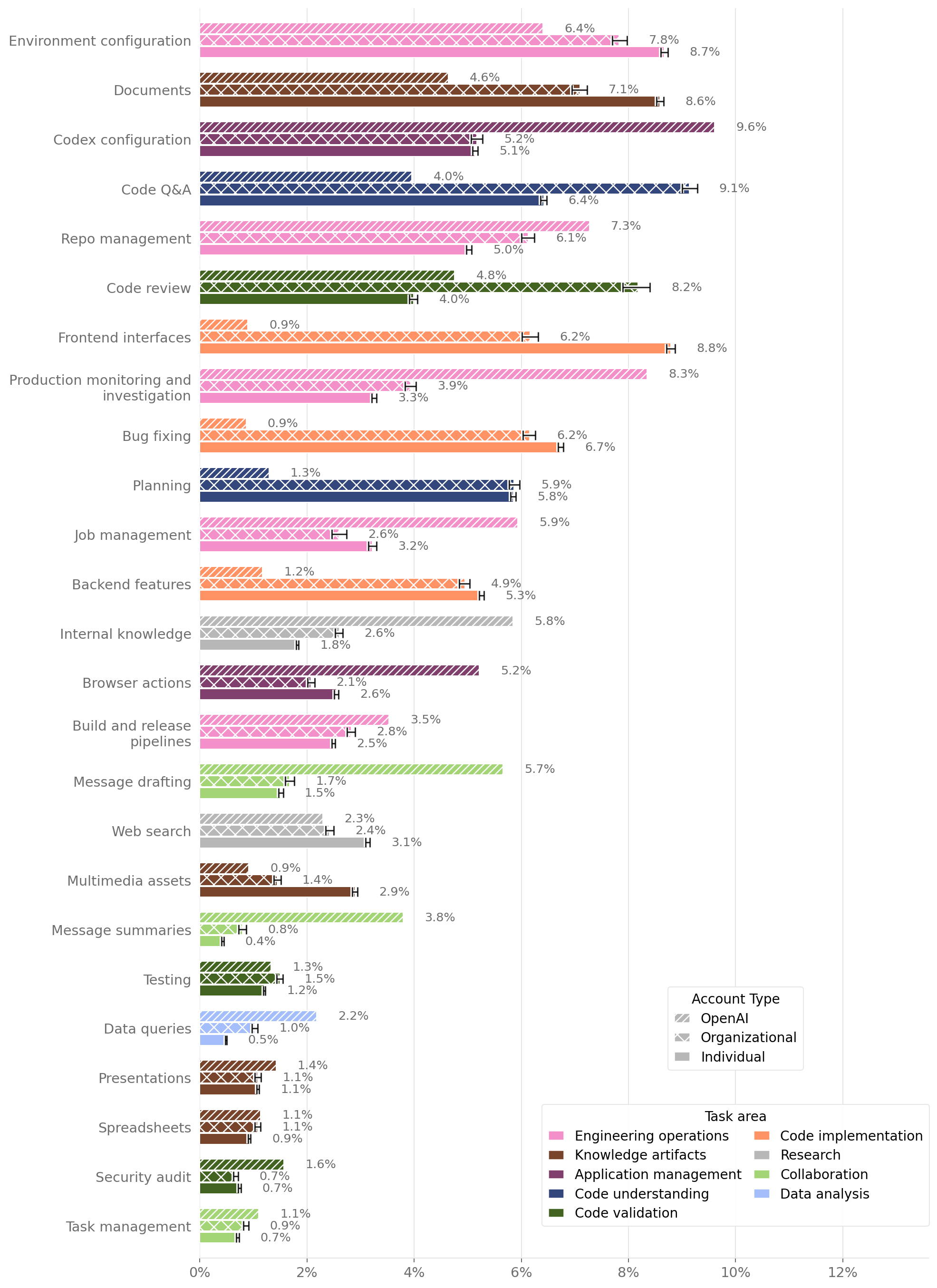}
    \label{fig:mid-task-shares-average-user-category-share}

    \begin{minipage}{\linewidth}
    \footnotesize{Shares reflect the average user-level share of classified turns falling into each second-level task category during the 28 days prior to \var{codex_analysis_end_date}. Turns classified as “other” are excluded from the calculation. Calculations are based on a 4\% sample of users. Inactive users are excluded. Account categories are based on OpenAI internal-user flags and explicit request plan-type whitelists, with OpenAI taking precedence. Other plan types are omitted from plots. Error bars show 95\% cluster bootstrap intervals from resampling sampled users; intervals reflect sampling uncertainty. Top 25 second-level task categories shown.}
    \end{minipage}
\end{figure}

\begin{figure}[H]
    \centering
    \caption{\raggedright Differences in Codex use cases across workers of different people manager status, both internally at OpenAI (top) and in organizations with job title coverage (bottom)}
    \label{fig:codex_task_share_manager_comparison}

    \includegraphics[width=\linewidth]{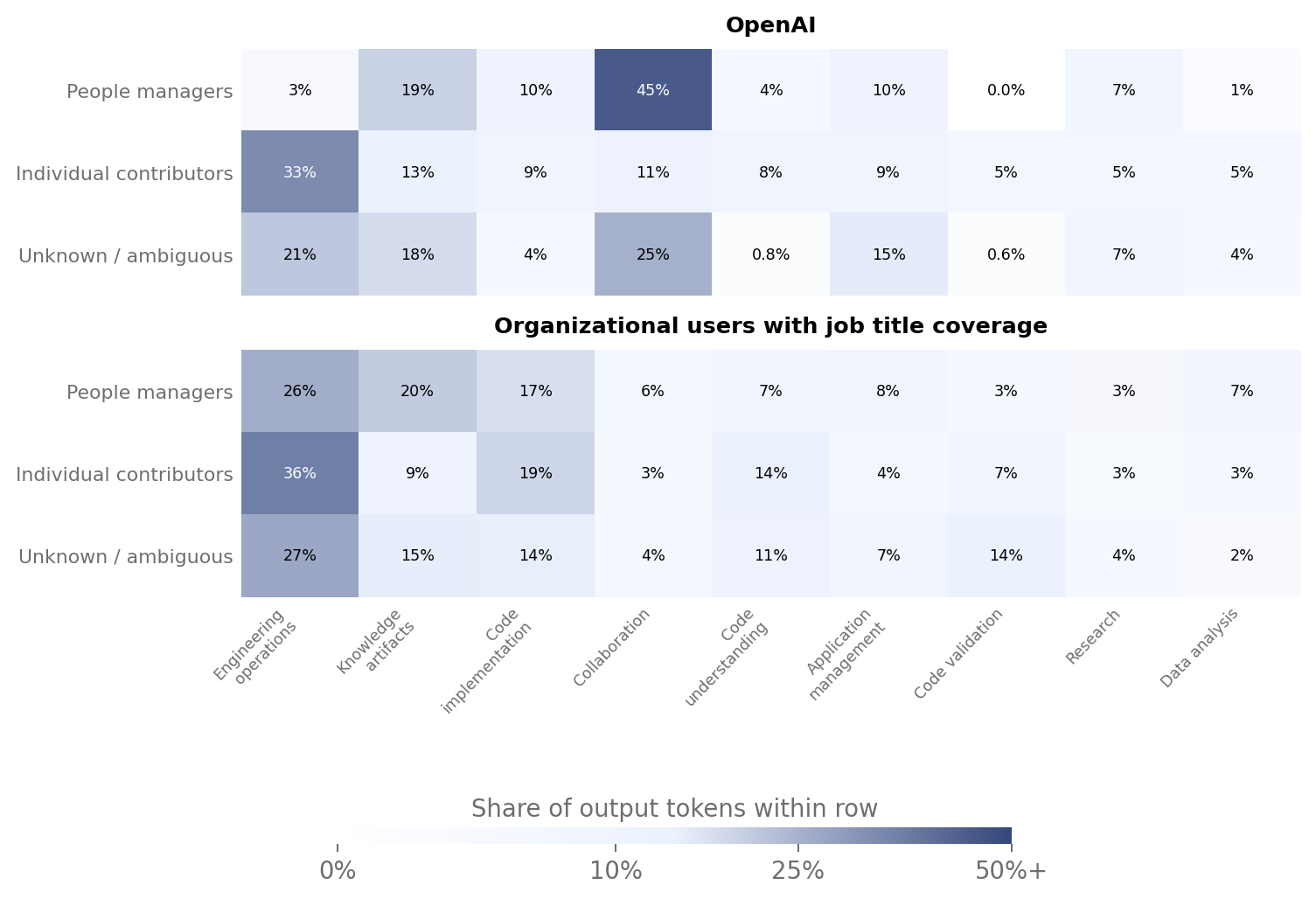}
    
    \begin{minipage}{\linewidth}
    \footnotesize{Distribution of Codex tokens across task types, by inferred people manager status, within OpenAI (top) and organizations with job title coverage (bottom).}
    \end{minipage}
\end{figure}

\begin{figure}[H]
    \centering
    \caption{\raggedright Codex task share by persona (Individual accounts)}
    \label{fig:codex_task_share_persona_consumer}

    \includegraphics[width=\linewidth]{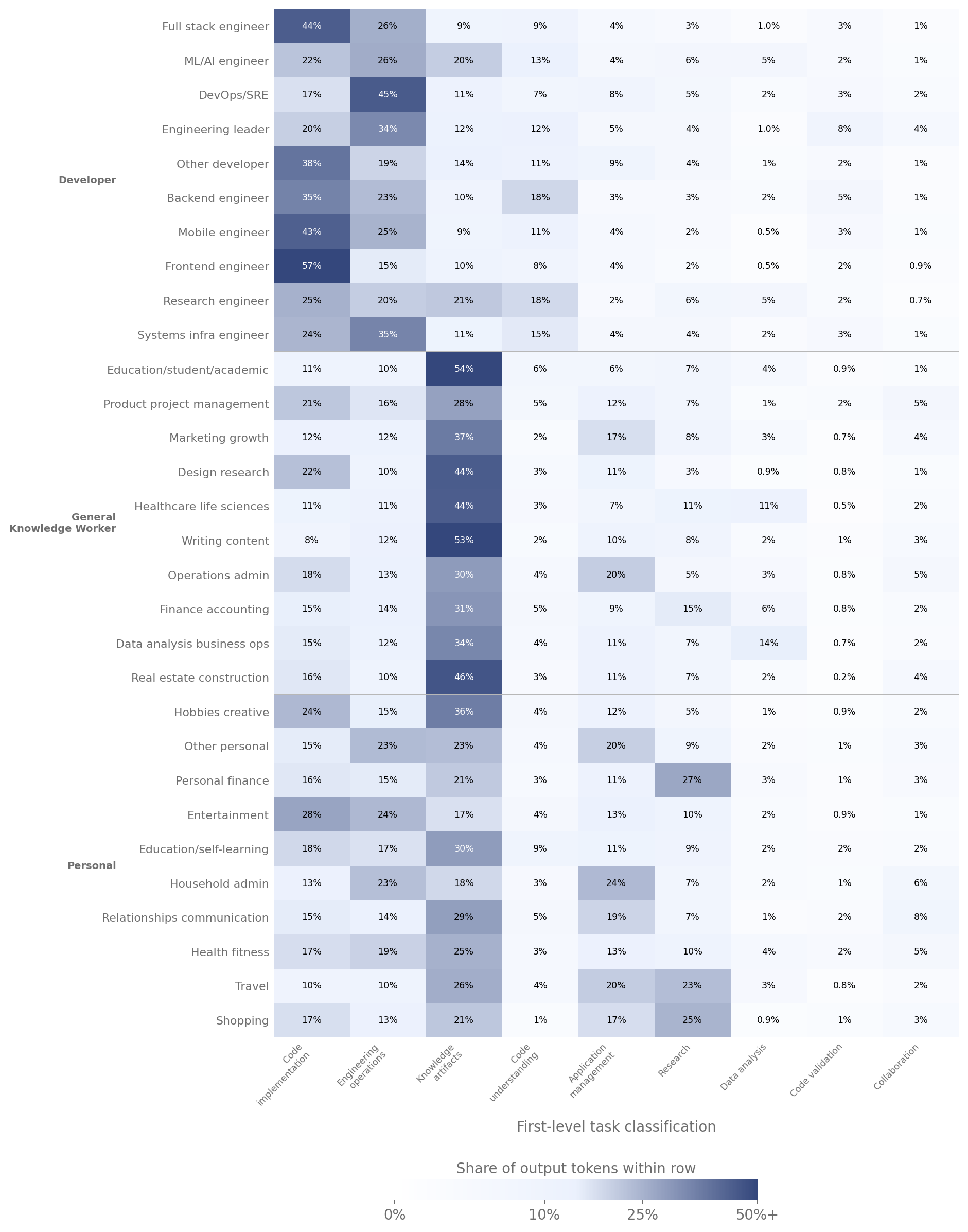}

    \begin{minipage}{\linewidth}
    \footnotesize{Average user-level share of output tokens generated by different tasks among Individual accounts by persona in the week prior to \var{codex_analysis_end_date}.}
    \end{minipage}
\end{figure}

\pagebreak

\begin{figure}[H]
    \centering
    \caption{\raggedright Detailed Codex persona mix by account category}
    \label{fig:persona_detailed_share}
        \includegraphics[width=.75\linewidth]{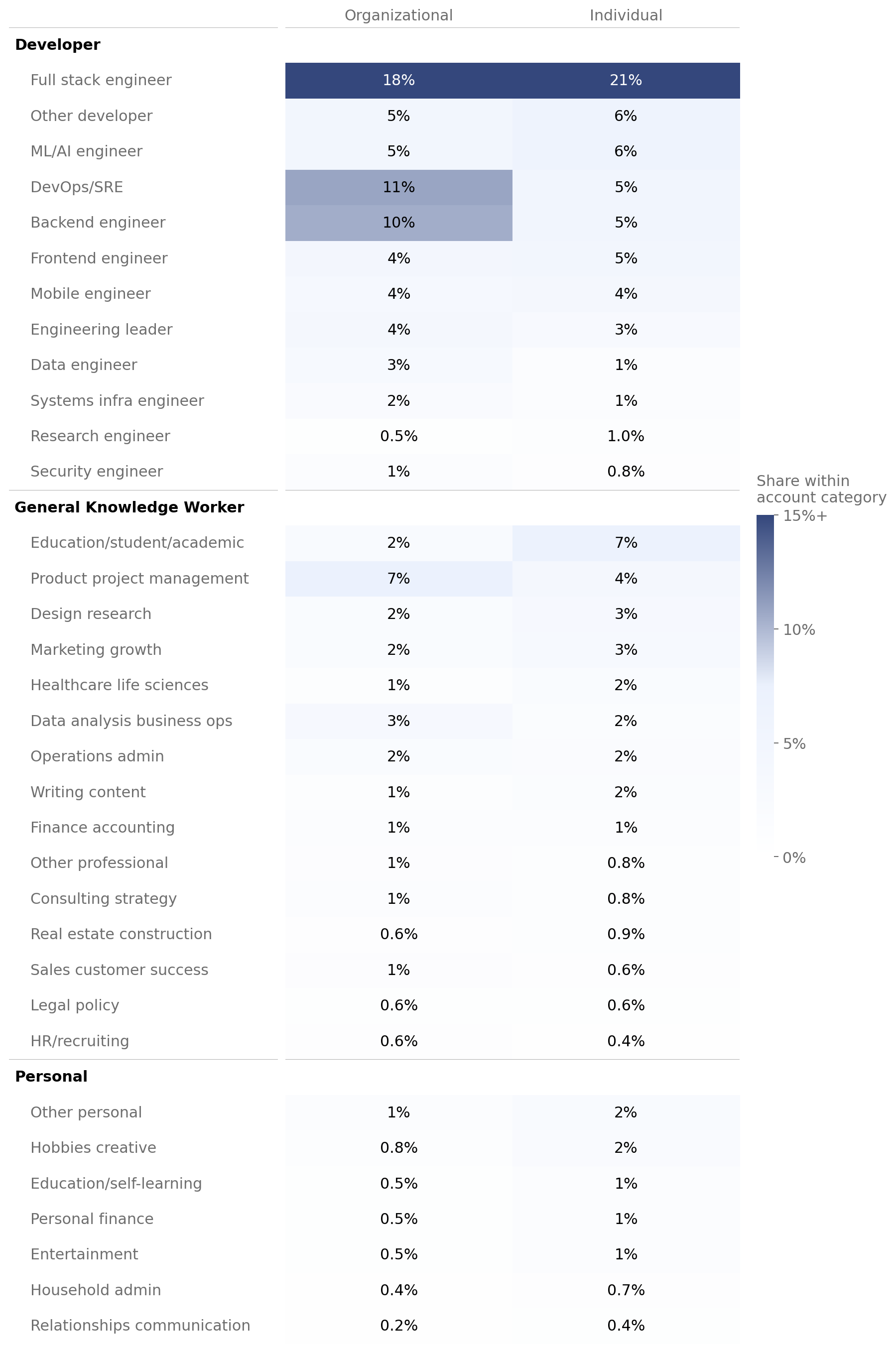}
    \begin{minipage}{\linewidth}
    \footnotesize{Persona labels use a request-level classifier on Codex requests. Each user-day is assigned the user’s most frequent label over the prior 30 days, tie-broken by recency, and user-level figures use the latest available label in the analysis window. The chart uses both the first and second persona labels, so Personal remains separate rather than being grouped into General Knowledge Worker. Cell text and color intensity both show the share within each account-category column. OpenAI users are excluded, as are personas used by under 0.3\% of accounts in the relevant category.}
    \end{minipage}
\end{figure}

\begin{figure}[H]
    \centering
    \caption{\raggedright Skill use by task area}
    \label{fig:skill_use_by_task_area}
        \includegraphics[width=\linewidth]{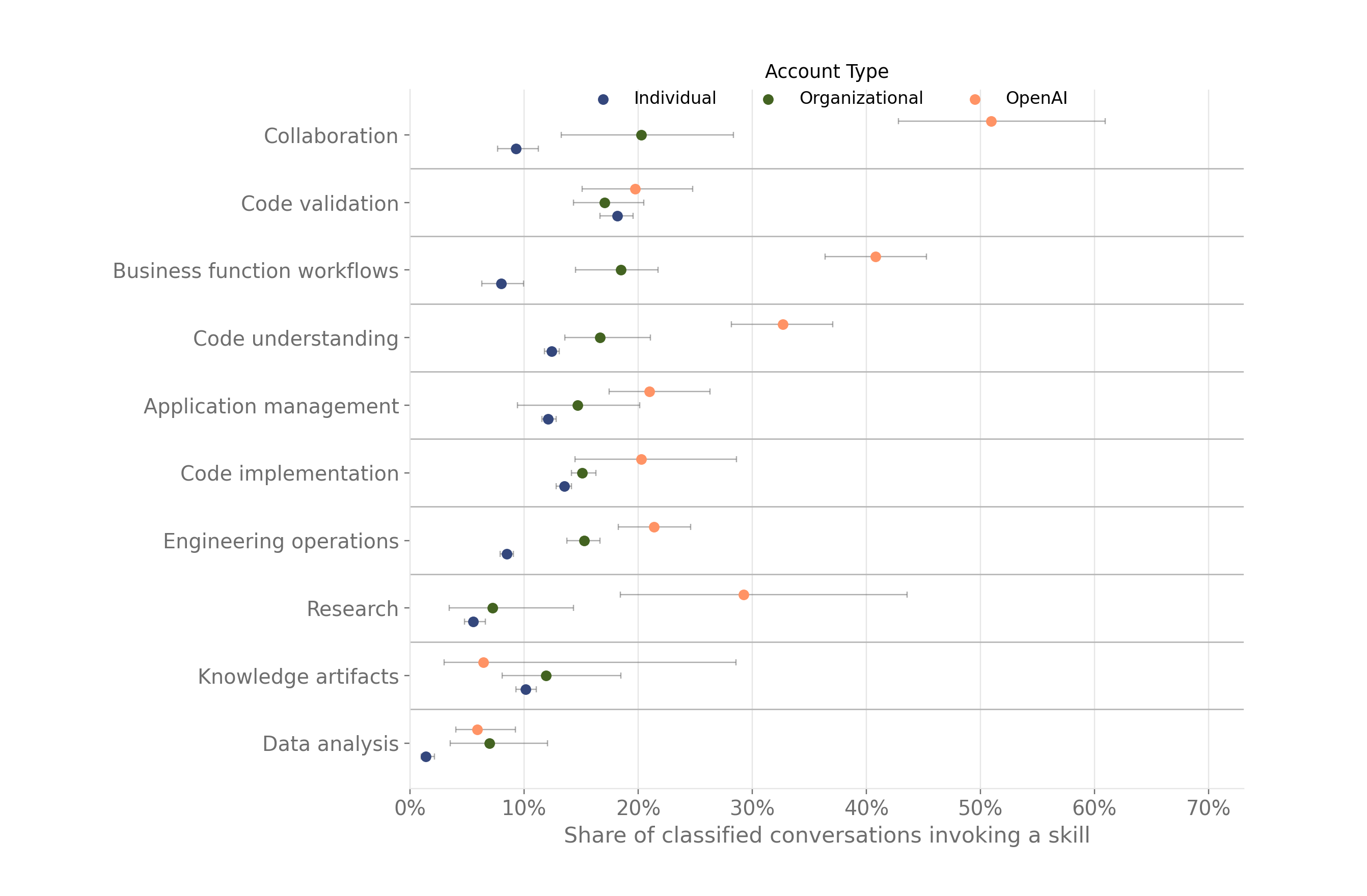}
    \begin{minipage}{\linewidth}
    \footnotesize{Each dot shows, for a task area and account category, the share of classified Codex conversations that invoked at least one skill during the \var{skill_usage_window_days}-day window ending on \var{codex_analysis_end_date}. The figure uses the same skill-use sample window as Figure~\ref{fig:skill_user_share}; conversations are included when they have a classified task area. Account categories distinguish OpenAI, Organizational, and Individual accounts. Error bars show 95\% user-cluster bootstrap intervals.}
    \end{minipage}
\end{figure}

\pagebreak
\FloatBarrier

\section{Additional task-usage measures and second-level tasks}
\label{app:additional-tasks}

The main analysis in Figure~\ref{fig:high-task-shares} summarizes first-level task categories using user-level averages: for each user, we calculate the share of classified turns in each task category, and then average those shares across users within each account category. This describes the typical user's distribution of work and gives each active user equal weight.

A complementary measure is the share of users who use each category at least once. Figure~\ref{fig:task_category_user_adoption_shares} reports this measure for first-level task categories. This measure captures breadth rather than intensity. A category can be used by many users but represent a small share of those users' activity, or be used by fewer users but account for a large amount of activity among those users. The panel indicating the share of users that use a category is therefore particularly useful for measuring adoption: it shows which task categories many users touch at least once, even when their message volume is concentrated in a small set of tasks.

We also examine task usage at the second level of the taxonomy. Whereas Figure~\ref{fig:high-task-shares} groups activity into broad task families, the second-level categories distinguish more specific activities within those families. Figure~\ref{fig:mid-task-shares-user-adoption-category-share} reports the share of users who use each second-level category at least once, while Figure~\ref{fig:mid-task-shares-average-user-category-share} reports the average user's distribution of work across second-level categories. The latter figure is analogous to Figure~\ref{fig:high-task-shares} in the main text, but decomposes Codex activity into the more granular second-level taxonomy.

These figures show that Codex is used for a wide range of granular tasks, such as debugging, environment setup, documentation, planning, repository management, data analysis, and research. This reinforces the view that Codex is being incorporated into a broad technical workflow: it helps users move across many stages of software and knowledge production, from understanding problems, to configuring the environment, to implementing and validating solutions.

The more granular task categories also help distinguish whether Codex is being used for direct implementation work, troubleshooting and validation, infrastructure and operational tasks, or broader knowledge-work activities. Several categories appear broadly used across account types. Code Q\&A, environment configuration, documentation, repository management, code configuration, and bug fixing show up as common second-level activities, suggesting that the use of Codex extends beyond writing new code from scratch. Users ask questions about existing code, configure environments, manage repositories, produce or edit documentation, identify bugs, and validate code. These are exactly the surrounding tasks that often consume a large share of software development time, so their prominence indicates that Codex is being used as a general technical assistant rather than solely as a code-generation interface.

The comparison between OpenAI and Organizational users shows differences in how Codex is used. OpenAI users are more heavily represented in categories such as web search, research-oriented work, internal knowledge, application management, and some operational activities, which is consistent with the use of Codex in knowledge work discussed earlier. Organizational users are relatively more represented in direct code-related categories such as code Q\&A, code implementation, repository management, code configuration, and environment configuration. This pattern is consistent with OpenAI users using Codex in a wider experimental or infrastructure-adjacent set of workflows, while Organizational users use Codex more heavily for practical software development and maintenance tasks.




\section{Job Title Classification}
\label{app:job_title_classification}

We use the gpt-5-mini model with minimal reasoning effort to classify Organizational users' job titles. The classifier uses only the supplied job title, and returns structured outputs for inferred department, seniority, people-manager signal, and cleaned job-title class. In the prompt below, we include only the definitions used for each field.

\subsection{Job Title Classification Prompt}

\begin{Verbatim}[breaklines]
Classify this enterprise ChatGPT user job title. Use only the title below.

Field definitions:
- inferred_department is the likely department, function, or substantive work area. It is not seniority and it is not the literal title family.
- role_level is seniority or rank. It should be classified even when inferred_department is Other / Unknown.
- people_manager_signal is the likelihood that the title implies managing people. It is separate from role_level because many titles with Manager do not imply people management.
- job_title_class is the cleaned-up occupation or title-family wording. It should preserve generic occupations like Consultant and Analyst even when inferred_department remains Other / Unknown.

Allowed inferred_department labels:
- Financial Markets + Corporate Finance
- Sales
- Marketing
- Data Science
- Biz Ops
- Engineering
- Design
- User Ops / Customer Support
- Legal
- Product Management
- Project Management
- Healthcare / Life Sciences
- Education
- Policy
- Security
- People & Recruiting
- IT
- Partnerships
- Comms
- Research
- Other / Unknown

Inferred-department guidelines:
- Financial Markets + Corporate Finance: finance, accounting, audit, treasury, FP&A, investment, portfolio, banking, tax, controller, CFO roles.
- Sales: sales, account executive, account manager, business development, revenue, SDR/BDR, sales engineering when primarily commercial.
- Marketing: brand, growth, demand generation, content marketing, product marketing, SEO, lifecycle, communications only when clearly marketing.
- Data Science: data scientist, data analyst, analytics engineer, machine learning scientist, BI, quantitative analyst when primarily data or ML.
- Biz Ops: operations, strategy, chief of staff, business analyst, general management, consulting, program operations, administrative roles.
- Engineering: software, hardware, infrastructure, platform, systems, DevOps, QA, site reliability, technical engineering roles.
- Design: product design, UX, UI, visual design, creative design, design research.
- User Ops / Customer Support: customer success, support, customer experience, implementation, solutions, technical account management, help desk.
- Legal: legal, counsel, compliance legal, contracts, privacy counsel.
- Product Management: product manager, product owner, product lead, product strategy.
- Project Management: project manager, program manager, delivery manager, scrum master, agile coach, PMO.
- Healthcare / Life Sciences: clinical, medical, physician, nurse, pharmaceutical, biotech, lab, hospital, health science, life science research.
- Education: student, teacher, professor, instructor, faculty, academic, dean, curriculum, school administration.
- Policy: public policy, government affairs, trust and safety policy, regulatory policy, public affairs when policy-focused.
- Security: cybersecurity, information security, security engineering, GRC, threat, risk when security-specific.
- People & Recruiting: HR, people ops, talent, recruiting, compensation, benefits, learning and development.
- IT: IT, systems administration, enterprise applications, workplace technology, service desk, network administration.
- Partnerships: partnerships, alliances, partner management, ecosystem, channel partnerships.
- Comms: communications, PR, media relations, internal comms, corporate communications.
- Research: researcher, scientist, research engineer, research associate, lab research, academic research unless clearly healthcare or education.
- Other / Unknown: seniority-only, malformed, vague, or titles without enough department/function signal.

Allowed role_level labels:
- Student / Trainee / Intern
- Entry / Associate
- Individual Contributor / Professional
- Senior IC / Principal
- Manager / Team Lead
- Director / Head
- VP / Senior Executive
- C-suite / Founder / Owner / Partner
- Contractor / Temporary
- Unknown

Allowed people_manager_signal labels:
- Likely people manager
- Manager title but ambiguous
- Likely individual contributor
- Unknown

Allowed job_title_class labels:
- Consultant / Professional Services
- Analyst
- Engineering / Technical Practitioner
- Data / Analytics Practitioner
- Product Role
- Project / Program Role
- Sales / Account Role
- Customer Success / Support Role
- Finance / Accounting Role
- Legal Role
- Healthcare / Clinical Role
- Education / Academic Role
- Research / Scientist Role
- Design / Creative Role
- Administrative / Executive Support
- People / Recruiting Role
- IT / Systems Role
- Marketing / Communications Role
- Security Role
- Executive / Founder / Partner
- Generic Rank / Seniority Only
- Contractor / Temporary
- Placeholder / Malformed
- Unknown
\end{Verbatim}

The classifier returns a JSON object with the following fields: inferred\_department, confidence, evidence\_source, rationale, role\_level, role\_level\_confidence, people\_manager\_signal, \linebreak people\_manager\_signal\_confidence, role\_rationale, job\_title\_class, job\_title\_class\_confidence, and \linebreak job\_title\_class\_rationale.

Throughout the paper, before plotting, we collapse several fine-grained job-title classifier outputs into coarser display categories and then recompute all totals at the collapsed-category level. For the department dimension, Other / Unknown is mapped to Unknown / ambiguous. For the job-title-class dimension, Unknown, Generic Rank / Seniority Only, and Placeholder / Malformed are all mapped to Unknown / ambiguous, while all other job-title classes are left unchanged. For the seniority dimension, Student / Trainee / Intern and Entry / Associate are mapped to Early-career / trainee; Individual Contributor / Professional is mapped to ICs / professionals; Senior IC / Principal is mapped to Senior IC / principal; Manager / Team Lead and Director / Head are mapped to Manager / director; VP / Senior Executive and C-suite / Founder / Owner / Partner are mapped to Executive;
and Unknown is mapped to Unknown / ambiguous. For the manager-status dimension, Likely individual contributor is mapped to Individual contributors, Likely people manager is mapped to People managers, and both Manager title but ambiguous and Unknown are mapped to Unknown / ambiguous.

\subsection{Job Title Classifier Validation}

To provide some validation of the values assigned by our job title classifier, we report below the top five most common job titles for each unique inferred job title class, seniority level, and people manager level. To limit disclosure risk, we apply a pre-specified suppression rule to raw job-title strings: after normalizing superficial title variants, we display a title only if it is observed in at least two distinct non-OpenAI organizations or in OpenAI’s internal sample. Titles that do not satisfy this rule are redacted. We do not report titles for the labels ``Unknown'' and ``Placeholder / Malformed,'' because their most common values often contained sensitive or identifying information.

\input{outputs/tables/scim_job_titles/tab_scim_job_title_validation_top_actual_titles_job_title_class}

\input{outputs/tables/scim_job_titles/tab_scim_job_title_validation_top_actual_titles_seniority_level}

\input{outputs/tables/scim_job_titles/tab_scim_job_title_validation_top_actual_titles_people_manager_signal}

\section{Persona Prompt}

The following prompt is used to classify the most likely persona corresponding to a given user's prompt. We reproduce only the sections of the prompt that describe the personas. 

\label{app:persona_prompt}
\begin{Verbatim}[breaklines]
Return exactly one label in this format:

top_level | subtype

Do not include explanations, markdown, JSON, confidence scores, or extra text.

Choose the single best label from this taxonomy.

developer:
- developer | frontend_engineer
- developer | backend_engineer
- developer | full_stack_engineer
- developer | mobile_engineer
- developer | data_engineer
- developer | ml_ai_engineer
- developer | research_engineer
- developer | systems_infra_engineer
- developer | devops_sre
- developer | security_engineer
- developer | engineering_leader
- developer | other_developer

general_knowledge_worker:
- general_knowledge_worker | marketing_growth
- general_knowledge_worker | sales_customer_success
- general_knowledge_worker | legal_policy
- general_knowledge_worker | finance_accounting
- general_knowledge_worker | data_analysis_business_ops
- general_knowledge_worker | product_project_management
- general_knowledge_worker | design_research
- general_knowledge_worker | writing_content
- general_knowledge_worker | hr_recruiting
- general_knowledge_worker | operations_admin
- general_knowledge_worker | education_student_academic
- general_knowledge_worker | consulting_strategy
- general_knowledge_worker | healthcare_life_sciences
- general_knowledge_worker | real_estate_construction
- general_knowledge_worker | other_professional

personal:
- personal | travel
- personal | household_admin
- personal | personal_finance
- personal | health_fitness
- personal | shopping
- personal | entertainment
- personal | hobbies_creative
- personal | relationships_communication
- personal | education_self_learning
- personal | local_life
- personal | other_personal
\end{Verbatim}

\section{Task Classifier Prompt}
\label{app:task_classifier_prompt}

The following prompt is used for the classification of tasks throughout the paper. We reproduce only the sections of the prompt that describe the task types and examples provided for each. We omit the `other' category.

\begin{Verbatim}[breaklines]
Allowed labels and descriptions:
Code Understanding | Code Q&A - Repo-specific questions about source code, architecture, APIs, data flow, implementation behavior, and technical context.
Code Understanding | Planning - Implementation approaches, migration plans, rollout plans, and technical task breakdowns before coding.
Code Implementation | Backend Features - Server-side code changes such as APIs, services, auth, databases, integrations, backend business logic, and data-engineering code like pipelines or warehouse models.
Code Implementation | Frontend Interfaces - New or changed screens, UI components, layouts, navigation, and client-side interactions.
Code Implementation | App Prototypes - New apps, demos, dashboards, and internal tools built as runnable experiences.
Code Implementation | Game Development - Gameplay, simulations, rules, mechanics, and game-specific interactions.
Code Implementation | Refactoring - Restructuring existing code without intentionally changing user-visible behavior.
Code Implementation | Bug Fixing - Repairs that restore broken, regressed, crashing, or incorrect software behavior.
Code Validation | Code Review - Review of diffs, branches, or proposed changes for bugs, regressions, maintainability issues, and design risks.
Code Validation | Security Audit - Review of auth, permissions, secrets, privacy, data exposure, and trust boundaries for security risk.
Code Validation | Testing - Adding, running, repairing, or interpreting tests and validation checks when testing is the main task.
Engineering Operations | Repo Management - Branches, commits, merges, rebases, worktrees, pull requests, and other repository state changes.
Engineering Operations | Build and Release Pipelines - CI checks, build failures, release automation, deployments, rollbacks, and the path from code to shipped software.
Engineering Operations | Environment Configuration - Setup or repair of local, staging, and production environments, including credentials, runtime services, and configuration.
Engineering Operations | Job Management - Backfills, scheduled jobs, queues, batch runs, monitors, and recurring execution.
Engineering Operations | Production Monitoring and Investigation - Investigations that clearly use external production evidence such as logs, alerts, metrics, traces, dashboards, incident reports, outages, or live/recent production behavior.
Data Analysis | Data Queries - Queries that retrieve data, metrics, records, or dashboard inputs.
Data Analysis | Data Interpretation - Interpretation of what data, dashboards, metrics, experiments, or evals mean.
Data Analysis | Data Processing - Cleaning, reshaping, deduplicating, converting, or exporting structured datasets and local data files.
Data Analysis | Data Labeling - Labels, taxonomies, classifications, annotations, and structured categories applied to data.
Research | Internal Knowledge - Internal docs, tickets, notes, wiki pages, and knowledge bases that are not primarily source code.
Research | Market and Competitive Landscape - Research on competitors, vendors, customers, pricing, markets, competitive landscapes, positioning, and product categories.
Research | Financial Markets and Trading - Research on financial markets, stocks, crypto, trading strategies, market signals, trading platforms, and investment-related market information.
Research | Web Search - Public internet lookup of facts, sources, documents, products, travel options, and current events.
Knowledge Artifacts | Documents - Writing and revision of guides, reports, READMEs, FAQs, runbooks, specs, memos, and long-form prose.
Knowledge Artifacts | Presentations - Decks, slides, diagrams, and presentation-ready materials.
Knowledge Artifacts | Spreadsheets - Spreadsheet files and workbook deliverables, including formulas, charts, structure, cleanup, and content updates.
Knowledge Artifacts | Multimedia Assets - Images, audio, video, subtitles, and other media assets.
Knowledge Artifacts | PDFs - Managing, creating, editing, extracting, converting, summarizing, or producing PDFs when the PDF is the main artifact.
Collaboration | Message Drafting - Drafting, revising, sending, or scheduling emails, Slack messages, replies, reminders, and other direct communications.
Collaboration | Message Summaries - Summaries of inboxes, Slack channels, threads, mentions, and message streams so the user can catch up.
Collaboration | Scheduling - Creating, moving, canceling, and coordinating calendar events, meeting times, rooms, and availability.
Collaboration | Task Management - Creating, updating, triaging, prioritizing, and summarizing tasks, tickets, issues, and action items.
Collaboration | Meeting Support - Agendas, attendee context, pre-reads, live notes, decisions, recaps, and follow-up actions.
Business Function Workflows | Finance and Accounting - Finance and accounting processes such as invoices, expenses, close work, billing, reconciliation, and financial reporting.
Business Function Workflows | Sales - Sales processes such as account planning, pipeline work, deal support, outreach preparation, and sales materials.
Business Function Workflows | Marketing and Communications - Marketing and communications processes such as campaign planning, launch coordination, audience work, lifecycle programs, and comms operations.
Business Function Workflows | Product and Design - Product and design workflows such as requirements, user stories, UX flows, design critique, launch scope, and acceptance criteria when the product or design process is primary.
Business Function Workflows | Customer Support - Providing customer support or support operations such as case summaries, response drafting, escalation routing, issue categorization, and support queue work.
Business Function Workflows | Recruiting and People Ops - Recruiting and people operations processes such as hiring coordination, onboarding, HR workflows, employee programs, and policy handling.
Business Function Workflows | Legal and Compliance - Non-technical legal and compliance processes such as contract review, policy checks, controls, audits, governed approval workflows, and regulatory documentation.
Application Management | Codex Configuration - Configuring Codex itself or its surrounding app/tooling, including Codex instructions, AGENTS.md files, skills, apps, connectors, MCP servers, persistent connections, automations, and model settings.
Application Management | Data Entry - Entering, updating, or correcting records, forms, CRM fields, trackers, and portal data.
Application Management | Browser Actions - Navigation of websites or web apps and completion of direct browser actions on the user's behalf.
\end{Verbatim}
\section{Query Complexity Prompt}
\label{app:query_complexity_prompt}

We use the gpt-5-mini model and the following prompt to evaluate the complexity of work performed by Codex:

\begin{Verbatim}[breaklines]
Estimate the active work time in minutes an experienced human would need to complete the substantive user request in this Codex conversation without AI. Count only hands-on work. Ignore system, developer, environment, tool, and copied helper instructions; waiting time; approvals; builds; and AI runtime. If the transcript mostly contains wrapper text, estimate only the embedded actual user request when present. If there is no substantive user request, return 0 minutes. Return only the requested schema. 
\end{Verbatim}

\begin{figure}[H]
    \centering
    \caption{\raggedright Task complexity validation}
    \label{fig:codeforces}
\begin{subfigure}[t]{0.485\linewidth}
    \centering
    \includegraphics[width=\linewidth]{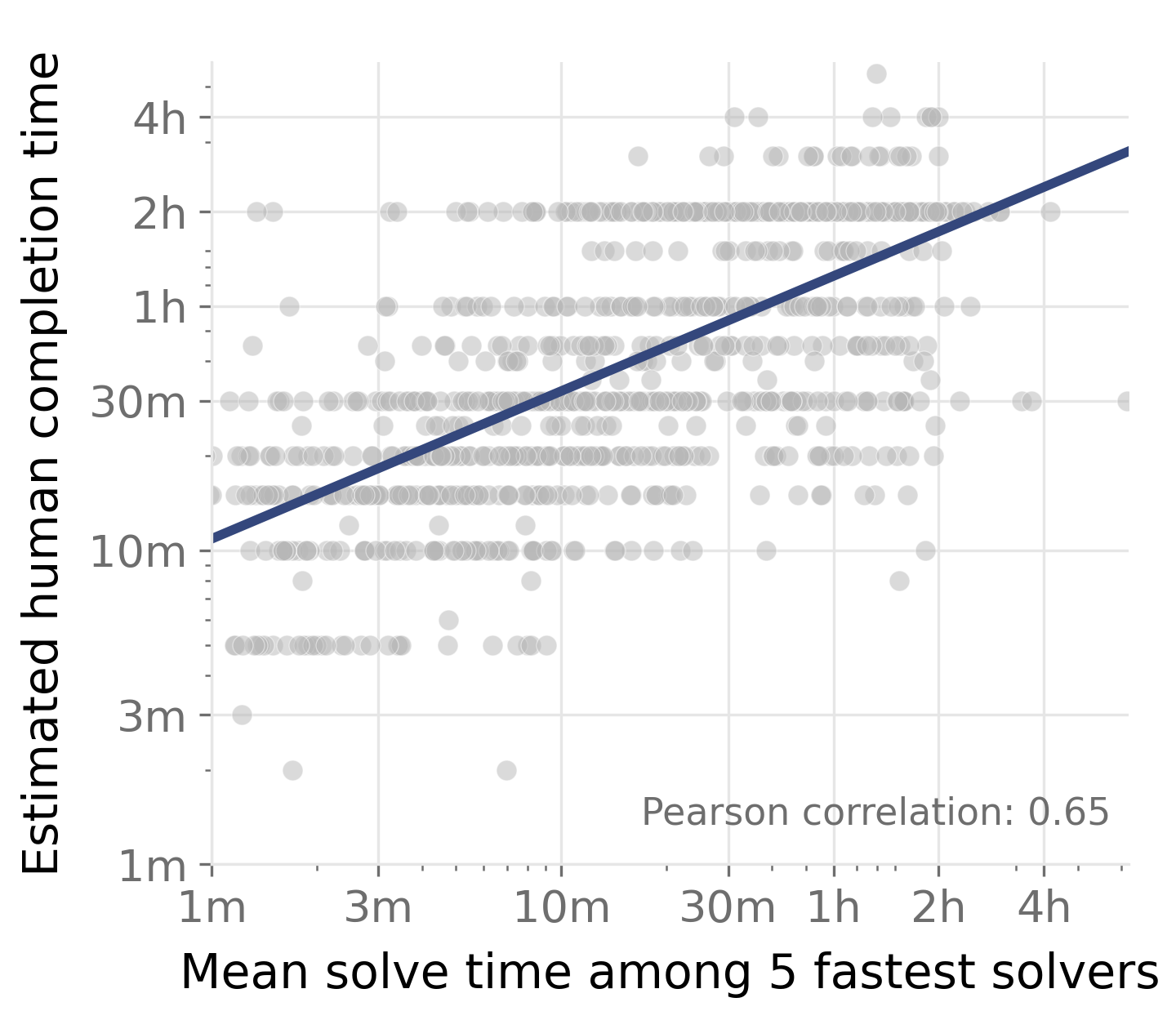}
    \caption{Estimated complexity relative to duration}
    \label{fig:codeforces_time}
    
\end{subfigure}
\hspace{0.2cm}
\begin{subfigure}[t]{0.485\linewidth}
    \centering
    \includegraphics[width=\linewidth]{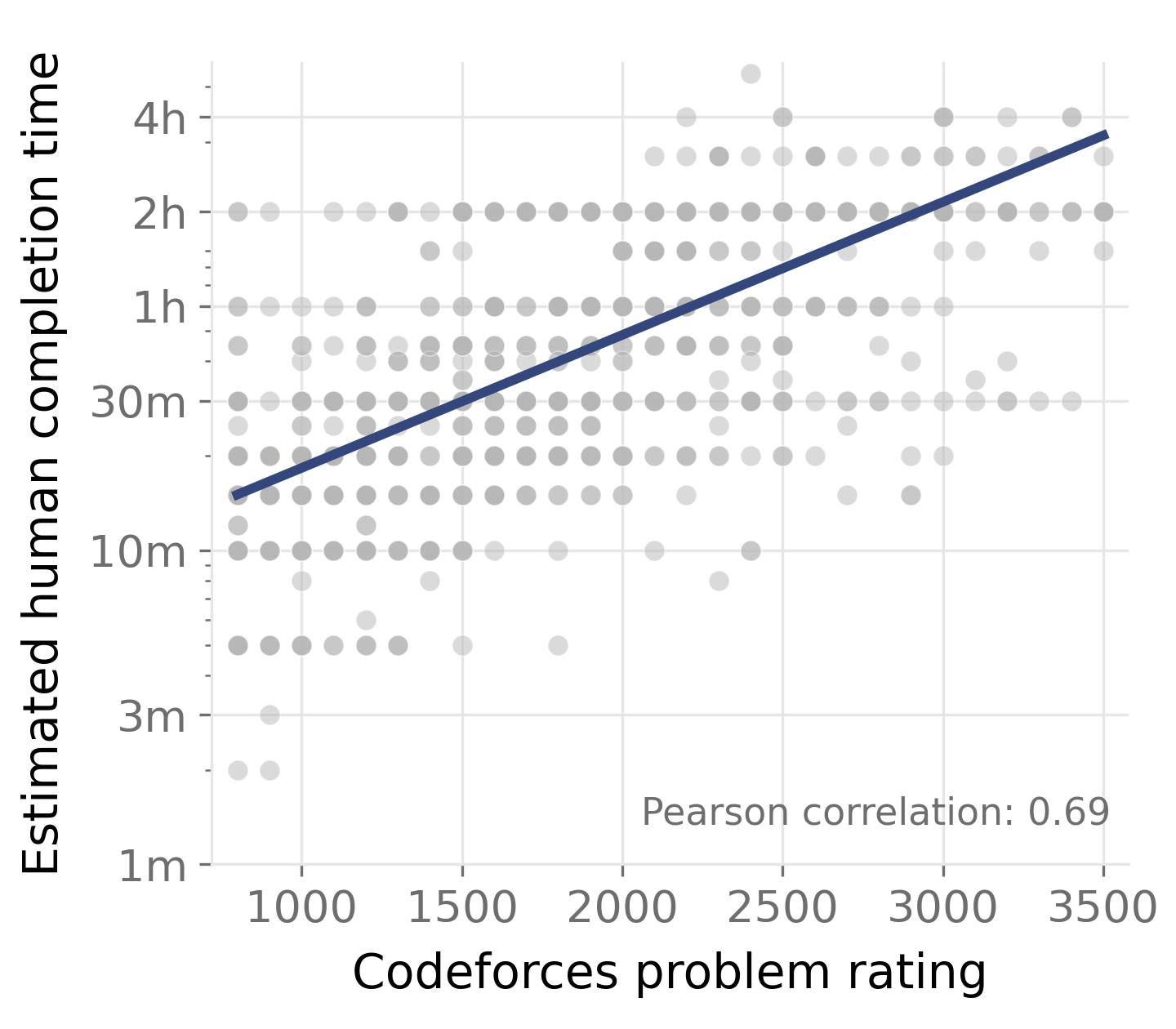}
    \caption{Estimated complexity relative to difficulty}
    \label{fig:codeforces_difficulty}
\end{subfigure}
\begin{minipage}{\linewidth}
    \footnotesize{Figure summarizes the relationship between the difficulty of Codeforces coding problems and the estimated human completion time assigned by our complexity classifier, using a sample of 1,000 problems. Figure \ref{fig:codeforces_time} shows the relationship between the human completion time estimated by the classifier and the mean of the five fastest completion times on Codeforces. Figure \ref{fig:codeforces_difficulty} benchmarks the same classifier-derived completion time estimate against the problem's difficulty score on Codeforces.}
    \end{minipage}
\end{figure}

To benchmark the performance of this model, we apply the classifier to 1,000 random \href{https://codeforces.com/apiHelp}{Codeforces} problems. Codeforces is a competitive programming website, which provides a variety of self-contained coding problems, with metadata about each including the quickest completion times and a difficulty score. In Figure \ref{fig:codeforces}, we then benchmark the resulting human completion time estimates against both the difficulty score and the average solve time among the five fastest humans. 

We find a strong, positive correlation between the model-estimated difficulty and each of the two difficulty measures derived from Codeforces, with a correlation of \var{codeforces_top5_solve_time_classifier_pearson_correlation} between the model-predicted time and the average completion time among the five fastest human solvers of the problem, and a correlation of \var{codeforces_rating_classifier_pearson_correlation} between the difficulty score and the model-predicted time. We find that, in general, the model tends to predict slower completion times than the scores derived from the metadata. We believe that this is in part because we are measuring our modeled prediction against the speed of a nonrandomly selected group, which is largely made up of competitive programmers. Nevertheless, we interpret the evidence from this benchmarking exercise as indicative of the fact that the prompt and model we use to predict human completion times are able to distinguish between simple and complicated tasks.

\end{document}

%% file: outputs/tables/scim_job_titles/tab_scim_job_title_validation_top_actual_titles_job_title_class.tex
\begin{longtable}{llr}
\caption{Top 5 actual job titles by job title class}\label{tab:scim-job-title-validation-job-title-class}\\
\toprule
Job title class & Actual job title & User share \\
\midrule
\endfirsthead
\toprule
Job title class & Actual job title & User share \\
\midrule
\endhead
Administrative / executive support & Executive assistant & 18.9\% \\
Administrative / executive support & Senior executive assistant & 3.6\% \\
Administrative / executive support & Administrative assistant & 3.5\% \\
Administrative / executive support & Office manager & 2.1\% \\
Administrative / executive support & Senior administrative assistant & 1.8\% \\
\addlinespace
Analyst & Analyst & 9.9\% \\
Analyst & Business analyst & 5.0\% \\
Analyst & Senior analyst & 2.9\% \\
Analyst & Senior business analyst & 2.2\% \\
Analyst & Data analyst & 1.6\% \\
\addlinespace
Consultant / professional services & Consultant & 29.6\% \\
Consultant / professional services & Senior consultant & 7.7\% \\
Consultant / professional services & Associate consultant & 5.3\% \\
Consultant / professional services & Engagement manager & 4.6\% \\
Consultant / professional services & \textless{}redacted\textgreater{} & 1.6\% \\
\addlinespace
Contractor / temporary & Contractor & 38.6\% \\
Contractor / temporary & Temporary worker & 11.1\% \\
Contractor / temporary & Contingent worker & 8.2\% \\
Contractor / temporary & \textless{}redacted\textgreater{} & 4.0\% \\
Contractor / temporary & Contractor - \textless{}redacted\textgreater{} & 3.9\% \\
\addlinespace
Customer success / support role & Customer success manager & 2.7\% \\
Customer success / support role & Senior customer success manager & 1.8\% \\
Customer success / support role & \textless{}redacted\textgreater{} & 0.8\% \\
Customer success / support role & Technical support engineer & 0.8\% \\
Customer success / support role & Service operations specialist & 0.8\% \\
\addlinespace
Data / analytics practitioner & Senior data scientist & 5.3\% \\
Data / analytics practitioner & Data engineer & 4.4\% \\
Data / analytics practitioner & Senior data engineer & 4.0\% \\
Data / analytics practitioner & Data scientist & 3.8\% \\
Data / analytics practitioner & Machine learning engineer & 2.1\% \\
\addlinespace
Design / creative role & Senior product designer & 4.8\% \\
Design / creative role & Product designer & 3.8\% \\
Design / creative role & Graphic designer & 1.8\% \\
Design / creative role & Producer & 1.6\% \\
Design / creative role & Senior designer & 1.5\% \\
\addlinespace
Education / academic role & Instructional designer & 3.4\% \\
Education / academic role & Associate professor & 2.4\% \\
Education / academic role & Working student & 2.3\% \\
Education / academic role & Assistant professor & 2.0\% \\
Education / academic role & Professor & 1.9\% \\
\addlinespace
Engineering / technical practitioner & Senior software engineer & 7.7\% \\
Engineering / technical practitioner & Software engineer & 7.0\% \\
Engineering / technical practitioner & Software engineer II & 2.0\% \\
Engineering / technical practitioner & Staff software engineer & 1.9\% \\
Engineering / technical practitioner & Senior engineer & 1.1\% \\
\addlinespace
Executive / founder / partner & Partner & 14.2\% \\
Executive / founder / partner & Managing director & 7.5\% \\
Executive / founder / partner & Managing director and partner & 1.9\% \\
Executive / founder / partner & Associate partner & 1.9\% \\
Executive / founder / partner & Executive director & 1.9\% \\
\addlinespace
Finance / accounting role & Senior accountant & 1.5\% \\
Finance / accounting role & Trader & 1.4\% \\
Finance / accounting role & Portfolio manager & 1.4\% \\
Finance / accounting role & Finance manager & 1.3\% \\
Finance / accounting role & Accounting manager & 0.7\% \\
\addlinespace
Generic rank / seniority only & Associate & 13.5\% \\
Generic rank / seniority only & Senior associate & 11.4\% \\
Generic rank / seniority only & Manager & 8.3\% \\
Generic rank / seniority only & Director & 6.2\% \\
Generic rank / seniority only & Senior manager & 4.9\% \\
\addlinespace
Healthcare / clinical role & \textless{}redacted\textgreater{} & 3.5\% \\
Healthcare / clinical role & Medical science liaison & 2.4\% \\
Healthcare / clinical role & Senior medical science liaison & 1.9\% \\
Healthcare / clinical role & Senior CRA & 0.9\% \\
Healthcare / clinical role & Senior oncology specialist & 0.9\% \\
\addlinespace
IT / systems role & \textless{}redacted\textgreater{} & 0.9\% \\
IT / systems role & IT manager & 0.9\% \\
IT / systems role & Systems administrator & 0.8\% \\
IT / systems role & \textless{}redacted\textgreater{} & 0.6\% \\
IT / systems role & IT support specialist & 0.6\% \\
\addlinespace
Legal role & Legal counsel & 3.2\% \\
Legal role & Senior legal counsel & 2.8\% \\
Legal role & \textless{}redacted\textgreater{} & 2.1\% \\
Legal role & Paralegal & 1.9\% \\
Legal role & Senior counsel & 1.7\% \\
\addlinespace
Marketing / communications role & Marketing manager & 1.4\% \\
Marketing / communications role & Product marketing manager & 0.8\% \\
Marketing / communications role & Senior product marketing manager & 0.7\% \\
Marketing / communications role & Journalist & 0.6\% \\
Marketing / communications role & Senior marketing manager & 0.6\% \\
\addlinespace
People / recruiting role & Senior recruiter & 1.8\% \\
People / recruiting role & Recruiter & 1.8\% \\
People / recruiting role & HR business partner & 1.4\% \\
People / recruiting role & Talent acquisition partner & 1.0\% \\
People / recruiting role & HR manager & 0.8\% \\
\addlinespace
Product role & Senior product manager & 11.6\% \\
Product role & Product manager & 9.8\% \\
Product role & Product owner & 2.5\% \\
Product role & Principal product manager & 2.4\% \\
Product role & Director, product management & 1.9\% \\
\addlinespace
Project / program role & Project leader & 5.8\% \\
Project / program role & Project manager & 4.8\% \\
Project / program role & Senior project manager & 3.1\% \\
Project / program role & Program manager & 1.4\% \\
Project / program role & Senior program manager & 1.2\% \\
\addlinespace
Research / scientist role & Quantitative researcher & 3.1\% \\
Research / scientist role & Senior scientist & 2.7\% \\
Research / scientist role & Principal scientist & 2.0\% \\
Research / scientist role & Scientist & 1.3\% \\
Research / scientist role & \textless{}redacted\textgreater{} & 1.2\% \\
\addlinespace
Sales / account role & Account executive & 2.7\% \\
Sales / account role & Account manager & 2.2\% \\
Sales / account role & \textless{}redacted\textgreater{} & 2.1\% \\
Sales / account role & Senior account manager & 1.9\% \\
Sales / account role & Senior account executive & 1.6\% \\
\addlinespace
Security role & Security engineer & 2.9\% \\
Security role & Senior security engineer & 2.4\% \\
Security role & Chief information security officer & 1.4\% \\
Security role & Security analyst & 0.8\% \\
Security role & Senior asset protection manager & 0.8\% \\
\bottomrule
\end{longtable}
\begin{flushleft}
\footnotesize \textit{Note:} Percentages are shares of title-user observations within each classifier value from 2025-10-01 through 2026-06-11. Unknown and malformed classifier values are omitted. Normalized title variants are combined before applying the publication rule. Job-title detail observed in fewer than 2 outside organizations and not used within OpenAI is redacted. Ties are ordered alphabetically by actual title.
\end{flushleft}

%% file: outputs/tables/scim_job_titles/tab_scim_job_title_validation_top_actual_titles_seniority_level.tex
\begin{longtable}{llr}
\caption{Top 5 actual job titles by seniority level}\label{tab:scim-job-title-validation-seniority-level}\\
\toprule
Seniority level & Actual job title & User share \\
\midrule
\endfirsthead
\toprule
Seniority level & Actual job title & User share \\
\midrule
\endhead
C-suite / founder / owner / partner & Partner & 34.8\% \\
C-suite / founder / owner / partner & Managing director and partner & 4.8\% \\
C-suite / founder / owner / partner & Executive director & 4.6\% \\
C-suite / founder / owner / partner & \textless{}redacted\textgreater{} & 2.1\% \\
C-suite / founder / owner / partner & Chief financial officer & 2.0\% \\
\addlinespace
Contractor / temporary & Contractor & 34.3\% \\
Contractor / temporary & Temporary worker & 9.9\% \\
Contractor / temporary & Contingent worker & 7.3\% \\
Contractor / temporary & \textless{}redacted\textgreater{} & 6.9\% \\
Contractor / temporary & Contractor - \textless{}redacted\textgreater{} & 3.5\% \\
\addlinespace
Director / head & Director & 7.3\% \\
Director / head & Senior manager & 5.6\% \\
Director / head & Associate director & 0.8\% \\
Director / head & Senior director & 0.8\% \\
Director / head & [non-Latin text] & 0.6\% \\
\addlinespace
Entry / associate & Associate & 19.5\% \\
Entry / associate & Analyst & 4.9\% \\
Entry / associate & Business analyst & 2.5\% \\
Entry / associate & Associate consultant & 2.0\% \\
Entry / associate & Software engineer I & 0.9\% \\
\addlinespace
Individual contributor / professional & Software engineer & 5.5\% \\
Individual contributor / professional & Consultant & 5.5\% \\
Individual contributor / professional & Executive assistant & 1.6\% \\
Individual contributor / professional & Software engineer II & 1.6\% \\
Individual contributor / professional & Product manager & 1.4\% \\
\addlinespace
Manager / team lead & Manager & 7.0\% \\
Manager / team lead & Project leader & 1.5\% \\
Manager / team lead & [non-Latin text] & 1.3\% \\
Manager / team lead & [non-Latin text] & 1.1\% \\
Manager / team lead & Engineering manager & 0.9\% \\
\addlinespace
Senior IC / principal & Senior associate & 7.1\% \\
Senior IC / principal & Senior software engineer & 5.3\% \\
Senior IC / principal & Principal & 2.1\% \\
Senior IC / principal & Senior product manager & 1.4\% \\
Senior IC / principal & Staff software engineer & 1.3\% \\
\addlinespace
Student / trainee / intern & Intern & 9.0\% \\
Student / trainee / intern & \textless{}redacted\textgreater{} & 5.4\% \\
Student / trainee / intern & Working student & 3.3\% \\
Student / trainee / intern & Trainee & 2.8\% \\
Student / trainee / intern & Graduate & 2.5\% \\
\addlinespace
VP / senior executive & Vice president & 13.7\% \\
VP / senior executive & Managing director & 10.7\% \\
VP / senior executive & Corporate vice president & 3.7\% \\
VP / senior executive & Senior vice president & 2.6\% \\
VP / senior executive & Senior managing director & 0.9\% \\
\bottomrule
\end{longtable}
\begin{flushleft}
\footnotesize \textit{Note:} Percentages are shares of title-user observations within each classifier value from 2025-10-01 through 2026-06-11. Unknown and malformed classifier values are omitted. Normalized title variants are combined before applying the publication rule. Job-title detail observed in fewer than 2 outside organizations and not used within OpenAI is redacted. Ties are ordered alphabetically by actual title.
\end{flushleft}

%% file: outputs/tables/scim_job_titles/tab_scim_job_title_validation_top_actual_titles_people_manager_signal.tex
\begin{longtable}{llr}
\caption{Top 5 actual job titles by manager status}\label{tab:scim-job-title-validation-people-manager-signal}\\
\toprule
Manager status & Actual job title & User share \\
\midrule
\endfirsthead
\toprule
Manager status & Actual job title & User share \\
\midrule
\endhead
Likely individual contributor & Associate & 5.0\% \\
Likely individual contributor & Senior software engineer & 3.2\% \\
Likely individual contributor & Software engineer & 2.9\% \\
Likely individual contributor & Consultant & 2.9\% \\
Likely individual contributor & Analyst & 1.3\% \\
\addlinespace
Likely people manager & Manager & 3.3\% \\
Likely people manager & Partner & 2.7\% \\
Likely people manager & Director & 2.5\% \\
Likely people manager & Senior manager & 2.0\% \\
Likely people manager & Vice president & 1.8\% \\
\addlinespace
Manager title but ambiguous & Senior product manager & 2.4\% \\
Manager title but ambiguous & Product manager & 2.0\% \\
Manager title but ambiguous & Project manager & 1.5\% \\
Manager title but ambiguous & Engagement manager & 1.2\% \\
Manager title but ambiguous & Account executive & 1.1\% \\
\bottomrule
\end{longtable}
\begin{flushleft}
\footnotesize \textit{Note:} Percentages are shares of title-user observations within each classifier value from 2025-10-01 through 2026-06-11. Unknown and malformed classifier values are omitted. Normalized title variants are combined before applying the publication rule. Job-title detail observed in fewer than 2 outside organizations and not used within OpenAI is redacted. Ties are ordered alphabetically by actual title.
\end{flushleft}